\documentclass[12pt]{article}
\usepackage{amsmath}
\usepackage{graphicx,psfrag,epsf}
\usepackage{enumerate}
\usepackage{natbib}
\usepackage{url} % not crucial - just used below for the URL
\usepackage[colorlinks,
            linkcolor=blue,
            anchorcolor=blue,
            citecolor=blue
            ]{hyperref}
\usepackage{float}
\usepackage{amsthm,amssymb}
\usepackage{multirow,booktabs}
\allowdisplaybreaks[4]

\newcommand{\blind}{1}

\newtheorem{theorem}{Theorem}[section]
\newtheorem{prop}{Proposition}[section]

\newtheorem{lemma}{Lemma}[section]
\newtheorem{remark}{Remark}[section]
\newtheorem{assumption}{Assumption}[section]

\newtheorem{MF}{Model Feature}[section]
\newtheorem{model}{Model}
\numberwithin{equation}{section}

\date{}
\begin{document}

\def\spacingset#1{\renewcommand{\baselinestretch}%
{#1}\small\normalsize} \spacingset{1}

\if1\blind
{
  \title{\bf Order Determination for Spiked Models}
  \author{Yicheng Zeng
    and
    Lixing Zhu\thanks{
    The authors gratefully acknowledge %\textit{
    the support from a grant from the University Grants Council of Hong Kong, Hong Kong, and an SNFC grant (NSFC11671042) from the National Natural Science Foundation of China. }\hspace{.2cm}\\
    Department of Mathematics, Hong Kong Baptist University, Hong Kong}
  \maketitle
} \fi

\if0\blind
{
  \bigskip
  \bigskip
  \bigskip
  \begin{center}
    {\LARGE\bf Order Determination for Large Dimensional Matrices}
\end{center}
  \medskip
} \fi
\bigskip

\begin{abstract}
Motivated by dimension reduction in regression analysis and signal detection, we investigate the order determination for large dimension matrices including spiked models of which the numbers of covariates are proportional to the sample sizes for different models. Because the asymptotic behaviour of the estimated eigenvalues of the corresponding matrices differ completely from those in fixed dimension scenarios, we then discuss the largest possible number we can identify and introduce a ``valley-cliff'' criterion. We propose two versions of the criterion: one based on the original differences of eigenvalues and the other based on the transformed differences, which reduces the effects of ridge selection in the former one. This generic method is very easy to implement and computationally inexpensive, and it can be applied to various matrices. As examples, we focus on spiked population models, spiked Fisher matrices and factor models with auto-covariance matrices. Numerical studies are conducted to examine the method's finite sample performances and to compare it with existing methods.
\end{abstract}

\vfill

\newpage
\spacingset{1.45} % DON'T change the spacing!
\section{Introduction}

Large dimensional matrices are often required to determine the order in diverse research fields to reduce the dimensionality. Examples include spiked population models proposed by \cite{johnstone2001distribution};
spiked Fisher matrices, which are motivated from signal detection and hypothesis testing for covariances; canonical correlation analysis; factor models; and target matrices in sufficient dimension reduction (see \cite{li1991sliced}; \cite{zhu2010sufficient}), which are for sufficient dimension reduction in regression analysis, in particular. \cite{luo2016combining} is a useful reference on order determination that proposed a ladle estimation for several models. We first use spiked population models as an example to describe the problem under study in this paper and propose a method that can be extended to handle other models.  For any spiked population model,  population covariance matrix $\Sigma_p$ can be written as a finite-rank perturbation of the identity matrix: $\Sigma_p=\sigma^2\textbf{I}_p+\Delta_p$, where $rank(\Delta_p)=q$ amounts to the fixed number of spikes, and $p$ is the dimension of the matrix.
Thus, determining the number of spikes is equivalent to determining the order of the matrix $\Delta$ mentioned above. For other large dimensional matrices, such as sample auto-covariance matrices and spiked Fisher matrices, the problems can be formulated in a similar manner.

The literature includes several proposals in the fixed dimension cases, such as the classic Akaike Information Criteria and Bayesian Information Criteria. Several methods have been developed for sufficient dimension reduction that can also be used in the models mentioned above. The methods include the sequential testing method (\cite{li1991sliced}), the BIC-type criterion (\cite{zhu2006sliced}), ridge ratio estimation (\cite{xia2015consistently}) and ladle estimation (\cite{luo2016combining}). Some of them can even handle cases with divergent dimension problems in which $p/n\to 0$ at certain rates.

However, when the dimension $p$ is proportional to the sample size $n$ where $p/n \to c$ for $0<c<\infty$, the problem becomes much more challenging. Thus, some efforts have been devoted to this problem with use of the large dimensional random matrix theory (see for example \cite{kritchman2008determining}; \cite{onatski2009testing}). Again, consider spiked population models. When $p/n\rightarrow c$ for a constant $c>0$, using the results derived by \cite{baik2006eigenvalues}, \cite{passemier2012determining} introduced a criterion that counts the number of differences between two consecutive eigenvalues below a predetermined positive constant threshold. However, when there are equal spikes, the corresponding differences could also be smaller than the threshold they designed, and the criterion could then very easily define a smaller estimator than the true number. %Invoking the results by \cite{bai2008central},
 \cite{passemier2014estimation} further modified this method to suit cases with multiple spikes. Underestimation however remains an issue when, say, there are three or more equal spikes. In addition to the problem caused by spike multiplicity, the dominating effect by a couple of the largest eigenvalues also results in underestimation. That is, when a couple of eigenvalues are very large and the other eigenvalues are too close to $\sigma^2$ and the differences between these small spikes would also be very small.
For the number of factors from a factor model for high-dimensional time series, \cite{li2017identifying} proposed a similar criterion to that of \cite{passemier2014estimation}.
%It could also inevitably encounter the underestimation problem caused by spike multiplicity.
For spiked Fisher matrices, \cite{wang2017extreme} used the classical  scree plot to determine the number of spikes when a threshold is selected in a delicate manner. The underestimation is still an issue. %Again, the criterion easily fails to detect equal and relatively weak signals.
We demonstrate this phenomenon in the numerical studies below. Relevant references include \cite{lam2012factor} and \cite{xia2015consistently}.

In this paper, we introduce a novel and generic criterion when the dimension $p$ is proportional to the sample size $n$. The criterion is based on the  eigenvalue difference-based ridge ratios with the following features. First, the criterion can handle  spike multiplicity problem and alleviates the large eigenvalue dominance problem. Second, the criterion has a nice ``valley-cliff'' pattern such that the consistent estimator is at the ``valley bottom'' facing the ``cliff'' upon which all the next ratios exceed a threshold. %We therefore call it the valley-cliff estimator.
%This estimator is then not necessarily the global minimizer over all indices for which the classical methods search.
%The idea would be useful for criterion construction for other order determination problems.
Third, adding ridge values plays a very important role to make the ratios  more stable and creates the ``valley-cliff" pattern. Fourth, to reduce the sensitivity of the criterion to ridge selection, we suggest another version that uses transformed eigenvalues. Fifth, we also discuss in detail reducing the effect of model scale in the construction. The new method is also very efficient in computation.

The remainder of this paper is organised as follows. In Section \ref{sec2}, we propose a VAlley-CLiff Estimation (VACLE) and provide an optimal lower bound to show what order can be identified.  In Section~\ref{sec3}, we first note that the VACLE could be improved when we use a transformation-based valley-cliff estimation (TVACLE) to alleviate the criterion's sensitivity to the designed ridge value. In this section, we also  discuss in detail the methods to select transformation. In Section~\ref{sec4}, we give spiked population models, factor models with auto-covariance matrices and spiked Fisher matrices as applications. Section~\ref{sec5} contains numerical studies and compares the VACLE and the TVACLE with existing competitors. A real data example is analysed in Section~\ref{sec6}. Some concluding remarks are in Section~\ref{sec7}, and the proofs of the theoretical results are contained in the supplementary materials.

\section{Criterion construction and properties}\label{sec2}
In this section, we describe our motivation in detail and provide the construction steps and its properties.

\subsection{ Motivation } 
Consider a simple spiked population model. For a $p\times p$ matrix $\Sigma_p=\sigma^2\textbf{I}_p+\Delta_p$ with the eigenvalues $\lambda_{1}\geq\cdots\geq \lambda_{q_1}> \lambda_{q_1+1}=\cdots =\lambda_{p}=\sigma^2$ where $q_1$ is a fixed number and the scale parameter $\sigma^2$ is either known or unknown. %Spiked population models satisfy this requirement. %To motivate our method,  consider a very simple case assuming that $\lambda_{1}>\lambda_{2}>\cdots >\cdots \lambda_{q_1}> \lambda_{q_1+1}=\cdots = \lambda_{p}=\sigma^2$. %and then $\tilde \lambda_{i}>0$ for $1\le i< q_1$.
%The spiked population matrix can be written as $\Sigma_p=\sigma^2\textbf{I}_p+\Delta_p$.
Let $\tilde\lambda_{i}$ be the eigenvalues of $\Delta_p$ and then $\lambda_{i}=\tilde \lambda_{i}+\sigma^2$, $1\le i\le p$,
 %$\sigma^2+\tilde \lambda_{1} \ge \sigma^2+\tilde\lambda_{2} \ge  \cdots \ge \sigma^2+\tilde\lambda_{q_1}> \sigma^2+\tilde \lambda_{q_1+1}=\cdots = \sigma^2+\tilde\lambda_{p}=\sigma^2$
with  $\tilde \lambda_{1} \ge \cdots \ge  \tilde\lambda_{q_1}> \tilde \lambda_{q_1+1}=\cdots = \tilde\lambda_{p}=0$.
When $p$ is proportional to $n$, estimation of $\lambda_i-\sigma^2$ is no longer consistent to $0$. % which is different from the fixed $p$ situation that sample counterpart of $\tilde\lambda_i-\sigma^2$ converges to 0 for $i\ge q_1$.
%When $\sigma^2$ is unknown and $p$ is proportional to $n$, accurate estimation of $\sigma^2$ with a rapid rate of convergence is difficult, although the literature includes some proposals such as \cite{ulfarsson2008dimension}, \cite{johnstone2009consistency} and \cite{passemier2012determining}.
Thus, we do not directly use either $\lambda_{i}$ or $\tilde \lambda_{i}$ but rather $\delta_{i}=\lambda_{i}- \lambda_{i+1}=\tilde \lambda_{i}- \tilde \lambda_{i+1}\ge 0$ for $i=1, \cdots, q_1$ and $=0$ for $i=q_1+1, \cdots, p-1$.
%This sequence of differences has the property that
%Note that $\delta_{n,i}$ are all non-negative this criterion widens the gap around the critical position $i=q$ by take $*/0$ type ratios, leading to that the structural dimension $q$ would be better separated at the sample level. Also, it involves a forward searching procedure of indexes, which would, to some degree, relieve the aforementioned underestimation problem. Inspired by the construction of VACLE, we develop an analogous but simplified criterion for our estimation problem. We name it {\bf VACLE} criterion as the abbreviation of {\bf Thresholding Ridge Ratio}.
%
Consider a sequence of ratios as
$r_{i}:=\delta_{i+1}/{\delta_{i}},\ 1\leq i\leq p-2$.
%According to assumption \ref{1}, the ratios
%\begin{equation}\label{2.3}
%r_{i}:=\dfrac{\delta_{i+1}}{\delta_{i}}, 1\leq i\leq p-2.
%\end{equation}
%This ratio is clearly scale-invariant as it is exactly equal to the ratio when we re-scale all eigenvalues as $\lambda_i/\sigma^2$. Therefore, we can simply regard $\sigma^2$ as $1$. %
These ratios are scale-invariant and can then have the following property, when $i\le q_1$:
%\vspace{-0.8em}
\begin{equation}\label{2.4}
r_{i}=\dfrac{\delta_{i+1}}{\delta_{i}}=\dfrac{\delta_{i+1}/\sigma^2}{\delta_{i}/\sigma^2}=\left\{\begin{array}{lcl}
 \ge 0,&&for\ i<q_1,\\=0,&&for\ i=q_1. %\\0/0=1,&&for\ q_1+1\leq i\leq p-2,
 \end{array}\right.
\end{equation}
For any $ q_1+1\leq i\leq p-2$, $r_{i}=0/0$ cannot be well defined because the values could vary dramatically and thus be instable.
%More seriously, because, unlike the original eigenvalues, the differences are no longer monotonic and  as a function of indices has neither a convexity nor a concavity pattern and may then have several local minima.
Due to the non-monotonicity of the $\delta_i$'s, some ratios $r_{i}$, even for $1\le i\le q_1$, could converge to either $*/0$, $0/*$, $0/0$ or $*/*$ respectively, where $*$ stands for a positive value that could differ for each appearance.
%This is a typical multi-minima problem.
This instability also occurs at the sample level. Thus, we cannot simply use this sequence of ratios to construct a criterion.
Taking these into consideration, we define a sequence of ridge type ratios:
\begin{equation}\label{2.3.1}
r^R_{i}:=\dfrac{\delta_{i+1}/\sigma^2+c_n}{\delta_{i}/\sigma^2+c_n}, 1\leq i\leq p-2.
\end{equation}
It is noticeable, in the construction of $r_i^R$, that we use $\delta_i/\sigma^2$ instead of $\delta_i$ in order to keep selection of $c_n$ independent of the scale parameter $\sigma^2$.
With appropriately selected $c_n\to 0$, these ratios have the following property:
%We will choose a sequence of positive, but sufficiently small constants $c_n$ such that these ratios have the following property:
\begin{equation*}\label{2.4.1}
r^R_{i}=\dfrac{\delta_{i+1}/\sigma^2+c_n}{\delta_{i}/\sigma^2+c_n}=\left\{\begin{array}{lcl}
 \ge 0,&&for\ i<q_1,\\=c_n/(\delta_{q_1}/\sigma^2+c_n)\to 0,&&for\ i=q_1, \\c_n/c_n=1,&&for\ q_1+1\leq i\leq p-2,
 \end{array}\right.
\end{equation*}
These ratios have a very useful  ``valley-cliff'' pattern,
%Note that $r^R_{i}$, with respect to $i$, is neither convex nor concave. Thus, $q_1$ could be one of the local minimizers of $ r^R_{i}$ over all indices $i$. The conventional method of using the global minimizer/maximizer to define an estimator of $q_1$ fails to work; however, the defined ratios follow the ``valley-cliff'' pattern
because $q_1$ should be the index of $ r^R_{q_1}\to 0$ at a ``valley bottom'' facing the ``cliff'' valued at $1$ of all next ratios $r^R_{i}$ for $i>q$. This nice pattern gives us a good opportunity to accurately identify $q_1$, although we will show later that in the setting in which $p$ is proportional to the sample size $n$, the identifiability of $q_1$ at the sample level remains a serious issue.

We also note that
%a minor trade-off of adding the ridge $c_n$ is that
the ratios %lose the scale-invariance property because they
depend on $\sigma^2$ and $c_n$.
%because the ridge $c_n$ is not scale-invariant and depends on the unknown $\sigma^2$.
Under the aforementioned scale transformation $\hat\lambda_{i}\rightarrowtail (\sigma^2)^{-1} \hat\lambda_{i}$, if $\hat\lambda_{i}$ are the estimated eigenvalues,
\begin{equation*}
\hat r_{i}^R=\dfrac{\hat \delta_{i+1}/\sigma^2+c_n}{\hat\delta_i/\sigma^2+c_n}=\dfrac{(\sigma^2)^{-1}\hat \lambda_{i+1}-(\sigma^2)^{-1}\hat\lambda_{i+2}+c_n}{(\sigma^2)^{-1}\hat \lambda_{i}-(\sigma^2)^{-1}\hat\lambda_{i+1}+c_n}=\dfrac{\hat \lambda_{i+1}-\hat\lambda_{i+2}+\sigma^2 c_n}{\hat \lambda_{i}-\hat\lambda_{i+1}+\sigma^2 c_n}. 
\end{equation*}
Later, however, we will show that the range of selecting $c_n$ can be rather wide, and thus the criterion is  not very seriously affected by this cost when  $\sigma^2$ is  estimated, which can be shown in the numerical studies we conduct later.
%When the estimation $\hat\sigma^2$ is either too low or too high, the performance should be influenced; thus, a reliable estimation $\hat\sigma^2$ to alleviate this problem is deserving of further study.
%Fortunately, the numerical studies we conduct and report in this paper show that it is not a very serious issue. Furthermore, our numerical studies that are not reported in this paper show that estimation with this method is much less affected than estimation  $\hat \sigma^2$ when compared with other existing methods direct use of the scale-invariant version $\hat\lambda_i/\hat \sigma^2$.
In addition, we have a brief discussion about the estimation of $\sigma^2$ in Section~\ref{sec5}.
%, and further study for efficient estimation of $\sigma^2$ is  out of our main focus in this paper.
%In contrast, it is still deserving of further study for efficient estimation of $\sigma^2$, but this is out of our main focus in this paper.

\subsection{ Valley-cliff criterion and estimation consistency}
Let $\textbf{T}_n$  be a target sample matrix of $\Sigma_p$ and $\hat \lambda_{1}\geq \cdots\geq \hat \lambda_{p}$ be its eigenvalues.
Here notations $\hat\lambda_i$ and $\hat\delta_i$ are related to the sample size $n$, although the $n$'s in  subscripts have been omitted for brevity.
  Define their sample versions $\hat{r}^R_{i}$ of $r_{i}^{R}$ in (\ref{2.3.1}) with $\hat{\delta}_{i}=\hat{\lambda}_{i}-\hat{\lambda}_{i+1}$  as
\begin{equation}
\hat r_{i}^R:=\dfrac{\hat\delta_{i+1}/\sigma^2+c_n}{\hat\delta_{i}/\sigma^2+c_n},\ 1\leq i \leq p-2,
\end{equation}
where $\sigma^2$ should be replaced by $\hat\sigma^2$ when $\sigma^2$ is unknown.

%\begin{equation}
%\hat{r}_{n,i}:=\dfrac{\hat{\delta}_{n,i+1}}{\hat{\delta}_{n,i}}
%=\dfrac{\hat{\lambda}_{n,i+1}-\hat{\lambda}_{n,i+2}}{\hat{\lambda}_{n,i}-\hat{\lambda}_{n,i+1}}, 1\leq i\leq p-2.
%\end{equation}
%Based on this sequence of ratios, we could construct a criterion from which a maximum index of the ratios smaller than a threshold value $0<\tau <1$ would be an estimator.
However, completely unlike the case with fixed $p$, even in the simple spiked population model case, $\hat{\lambda}_{i}$ are not  consistent to $\lambda_{i}$ and these ratios cannot then simply converge to those in (\ref{2.4}). The number $q_1$ is generally unidentifiable. In the following section, we give the largest possible order we can identify.
Define
\begin{equation}\label{q}
q:=\# \{i: 1\le i\le q_1, \,  \lambda_i>U(F)>\sigma^2 \}
\end{equation}
for some constant $U(F)$ where $F$ is the limiting spectral distribution (LSD) of all estimated eigenvalues $\hat \lambda_i$'s with the support $(a(F), b(F))$.
The constant $U(F)$ is the phase transition point (see \cite{baik2006eigenvalues}) and also the optimal bound for identifiability.
We still use a spiked population model as a typical example. From \cite{baik2005phase} and \cite{baik2006eigenvalues}, %established the well
%known BBP phase transition phenomenon for the largest eigenvalue of a complex Gaussian
%sample covariance matrix, and then a refinement without distribution assumption was stated in \cite{baik2006eigenvalues}. This phenomenon clarified a fact that a leading sample eigenvalue, for instance $\hat\lambda_1$, becomes an outlier away from the support of LSD if and only if its population counterpart $\lambda_1$  exceeds the  critical value $(1+\sqrt c)\sigma^2$. In other words,
any spike with strength not stronger than  $(1+\sqrt c)\sigma^2$ is not identifiable. In this case, $U(F)$ refers to the critical value $(1+\sqrt c)\sigma^2$.
%Section~\ref{sec4} shows that different models have have different LSDs, supports and $U(F)$. And the relationships among these three show different patterns in different models.
More details are included in Section~\ref{sec4}.

%Let $q:=\# \{ \lambda_i:\lambda_i>\sigma^2(1+\sqrt c) \}$ where $c=\lim_{n\to \infty}p/n$. Clearly, $q\le q_1$ and the equality holds if and only if $c=0$. In the proof of the following theorem, we will show that when $c>0$, it is impossible to identify $q_1$. That is, what we can do is just to identify $q$.

Selecting an appropriate sequence $c_n$ plays an important role for estimation efficiency.
%We will select it in the following principle:  $\hat{\delta}_{i}=o_p(c_n)$ for $q+1\le i \le p-1$. This selection can guarantee that under the conditions given below,
When it is selected in the principle:  $\hat{\delta}_{i}=o_p(c_n)$ for $q+1\le i \le p-1$, %the sequence of
$\hat r_{i}^{R}$ still have a nice ``valley-cliff'' pattern at $i=q$ as
\begin{equation}
\lim_{n\to+\infty}\hat r_{i}^{R}=\left\{\begin{array}{lcl}
0,&&i=q\\1,&&q+1\leq i\leq L-1
\end{array}  \right.
\end{equation}
where $L$ is a prefixed upper bound for $q$.
%From Bai and Yao (2008), $\hat\lambda_{n,i}$, for $1\leq i\leq q$, is $\sqrt n$-consistent, which implies that $\hat{\delta}_{n,i}=O_p(n^{-1/2})$ for $1\le i\le q$. Thus, the value of $\hat{r}_{n,i}^R$, for $i<q$, depends on the magnitude of ridge sequence $c_n$. However, it does not affect our following criterion.
Taking this advantage, we define a thresholding valley-cliff estimator as, for a constant $\tau$ with $0< \tau <1$
\begin{equation}\label{10}
\hat{q}_n^{VACLE}=\max_{1\leq i\leq L-2}\left\{i:\hat r_{i}^{R}\leq\tau\right\}.
\end{equation}
%\begin{equation}\label{10}
%\hat{q}_n^{VACLE}=\max_{1\leq i\leq L-2}\left\{i:\dfrac{\hat \lambda_{i+1}-\hat\lambda_{i+2}+\sigma^2 c_n}{\hat \lambda_{i}-\hat\lambda_{i+1}+\sigma^2 c_n}\leq\tau\right\}.
%\end{equation}

%\begin{remark}
%The upper bound $[\alpha \min\{n,p\}]-1$ for $q$ in the above estimator is guaranteed by these results about the convergence of $\hat{\lambda}_{n,i}$'s. More practically, we can simply use  a prefixed upper bound $L-2$:
%\begin{eqnarray}\label{11}
%\hat{q}_n^{VACLE}=\max_{1\leq i\leq L-2}\{i:\dfrac{\hat \lambda_{n,i+1}-\hat\lambda_{n,i+2}+c_n}{\hat \lambda_{n,i}-\hat\lambda_{n,i+1}+c_n}\leq\tau\}.
%\end{eqnarray}
To handle more general models, we consider the large dimensional matrices with the following model features.
\begin{MF}\label{1}
There exists a bound $U(F)$ such that the number $q$ defined in (\ref{q}) is a fixed constant and satisfies: \\
(A1) %$\exists\ q,L\in \mathbb{N^+}$ and
there is a value $d$ such that $\hat\lambda_{q}/\sigma^2- d=o_P(1)$ as\ $n\rightarrow \infty$;\\ %. Accordingly, $\hat\delta_{n,q}\xrightarrow{P} d-e>0$ and $\hat\delta_{n,i}\xrightarrow{P} 0$, for $q+1\leq i\leq L-1$;\\
(A2) for a large fixed value $L$ satisfying $q+1<L< p$, there is a constant $e<d$ and a sequence $\tilde c_n\rightarrow 0$ such that  $\hat\lambda_{i}/\sigma^2-e=O_p(\tilde c_n)$, for $q+1\le i\le L$.
\end{MF}
\begin{remark}
Model Feature \ref{1} describes features of the model structure at the sample level, that essentially requires
%the model to satisfy
certain assumptions at the population level. %Model Feature \ref{1} coincides with the typical features of finite rank perturbation of large dimensional random matrices that have
%limiting spectral distributions
% LSDs with compact supports.
Condition $(A1)$ corresponds to the so-called {\it phase transition phenomenon} for the extreme eigenvalues, and $(A2)$ further focuses on the fluctuations of those that stick to the boundary of the support of the LSD.
%limiting spectral distribution.
General theory about the phase transitions and fluctuations can be found, for example, in \cite{peche2006largest}, \cite{benaych2011fluctuations}, \cite{benaych2011eigenvalues} and \cite{knowles2017anisotropic}.  The details  about how these features can be exhibited in three types of models are given in Section~\ref{sec4}. %, which shows that different matrices will have different values $q$ that are identifiable and highly related to the LSD. % of the estimated eigenvalues.
%\cite{han2016tracy}, and \cite{han2018unified}.
  %Those theories show that we have no chance to identify the eigenvalues  between $\sigma^2$ and  the upper bound $U(F)$ of the support set of limiting spectral distribution. This is  because, as (A2) described, there are too many estimated eigenvalues converging to the same value even when the corresponding eigenvalues at the population level are different. Thus,
%Condition (A2) restricts $q$ to be bounded in order to ensure the consistency of our criterion, and an extension to the case of divergent $q$ will be discussed in our future work.
\end{remark}

The estimation consistency is stated as follows.
\begin{theorem}\label{2}
When a model satisfies Model Feature~\ref{1}, and $\tilde c_n=o( c_n)$, then $\mathbb{P} (\hat{q}_n^{VACLE}=q)\to 1$ as $n\to \infty$.
\end{theorem}

\begin{remark}
The convergence rate of $\hat\lambda_{i}$ to a constant $e$, for $q+1\leq i\leq L-1$ is often $O_p(n^{-2/3})$, namely $\tilde c_n=n^{-2/3}$. The references include \cite{benaych2011fluctuations}, \cite{bao2015universality}, \cite{han2016tracy} and \cite{han2018unified} for several models discussed in Section~\ref{sec4}. However, for a spiked auto-covariance matrix, we will also state our result provided, as \cite{li2017identifying} did, that this rate can be achieved as this rate has not yet formally been derived. In this paper, the ridge $c_n \to 0$ is  only restricted to  $\hat\lambda_{q+1}=o_p(c_n)$. Such a wide range for the ridge selection alleviates, to a great extent, the influence from $\sigma^2$ when it needs to be estimated. The estimation issue will be discussed in Section~\ref{sec5}.
%For the bound $U(F)$ of the significant eigenvalues, the analyses for the models in Section~\ref{sec4} show that the bound is optimal. That is, it is impossible in theory to identify any eigenvalues $\lambda_i$ for $i>q$ with $\sigma^2< \lambda_i< U(F)$. Furthermore, in Section~\ref{sec4}, we also consider the auto-covariance matrix for which the bound cannot be determined so simply, but an optimal bound can still be obtained.
\end{remark}

\section{Modification of the VACLE}\label{sec3}

Although Theorem~\ref{2} provides estimation consistency, some numerical studies
%\subsection{Motivation and Construction of TVACLE}
that are not presented in this paper indicate that the performance of $\hat{q}_n^{VACLE}$ is sometimes and somehow sensitive to the choice of the ridge $c_n$ in finite sample cases.
%where $d$ and $e$ are defined in Model Feature~\ref{1}.
To be  specific, when $d-e$ in Model Feature~~\ref{1} is small, the ratio at $q$ could be close to $1$, and then we would  easily determine a smaller value. Therefore, a small ridge $c_n$ is in demand. On the other hand, a small $c_n$  would result in the instability caused by $0/0$ type ratios, and overestimation would be possible. Thus,  a trade-off exists between underestimation and overestimation in the choice of ridge $c_n$. We now attempt to alleviate this dilemma by using transformed eigenvalues.

%Recall that
%\begin{equation}
%\hat r_{n,i}^R=\dfrac{\hat\delta_{n,i+1}+c_n}{\hat\delta_{n,i}+c_n},\ 1\leq i\leq p-2,
%\end{equation}
%where $\hat\delta_{n,i}=\hat\lambda_{n,i}-\hat\lambda_{n,i+1}\geq0$. Note that $\hat\delta_{n,i}$ converges to a positive constant for $i=q$ but to zero for $q+1\leq i\leq L-2$. Also, recall the idea of the VACLE is the following.  The selected ridge $c_n$ dominates the values of $\hat \delta_{n,i}$ and then $\hat r_{n,i}^R$ when $q+1\leq i\leq L-2$, while $\hat\delta_{n,q}$ dominates $c_n$ and then $\hat r_{n,q}^R$. Thus, to further separate these ratios, we use transformed $\hat\delta_{n,i}$  to strengthen these dominations.
Considering a transformation (depending on $n$) $f_n(\cdot)$, define
\begin{equation}
\hat\delta_{i}^*=f_n(\hat\lambda_{i}/\sigma^2)-f_n(\hat\lambda_{i+1}/\sigma^2),\ i=1,2,\cdots,p-1.
\end{equation}
%where the transformation function sequence $\{f_n\}_{n\geq 1}$ will be discussed later. It is worth mentioning that we make the transformation on the initial eigenvalues $\hat\lambda_{n,i}$ instead of on the difference $\hat\delta_{n,i}$. Unlike some other methods, for example simply taking the square of $\hat\delta_{n,i}$ to raise its convergence rate, our transformation, to some extent, equates multiplying $\hat\delta_{n,i}$ by a constant, keeping its convergence rate remain the same. However, for the case of $e$ being close to $d$, the $\hat\delta_{n,q}$ would be a pretty small number comparable to $\hat\delta_{n,q+1}$, and then taking square of $\hat\delta_{n,i}$ would bring down the value of $\hat\delta_{n,q+1}$ as well as $\hat\delta_{n,q}$, making the gap between $\hat r_{n,q}^R$ and $\hat r_{n,q+1}^R$ still vague. For the performance of our transformed one in that case, we will discuss it in details in the following.
The ratios are defined as
\begin{equation}\label{ratio-TR}
\hat r_{i}^{TR}:=\dfrac{\hat\delta_{i+1}^*+c_n}{\hat\delta_{i}^*+c_n},\ 1\leq i\leq p-2.
\end{equation}
The estimator of $q$ is defined as
\begin{equation}
\hat{q}_n^{TVACLE}=\max_{1\leq i\leq L-2}\left\{i:\hat r_{i}^{TR}\leq \tau\right\},
\end{equation}
where $c_n$ and $\tau$ have the same definitions as before. We call this criterion the transformation-based valley-cliff estimation(TVACLE).

For any transformation $f_n$, we wish that $\hat r_{i}^{TR}$ remains close to $1$ for $i>q$, and $\hat r_{q}^{TR}$ is closer to zero than $\hat r_{q}^R$. To achieve these objectives, we consider a transformation that can satisfy the following requirements $(i)-(iii)$:\\
\indent (i) $\mathbb{P}\{\hat\delta_{q}^*\geq\hat\delta_{q}/\sigma^2\}\rightarrow 1 $;\\
\indent (ii) $\mathbb{P}\{\hat\delta_{i}^*\leq\hat\delta_{i}/\sigma^2\}\rightarrow 1,\ for\ q+1\leq i\leq p-2$;\\
\indent (iii) ${\hat\delta_{q+1}^*}/{\hat\delta_{q}^*}\leq {\hat\delta_{q+1}}/{\hat\delta_{q}}$.
%\begin{itemize}
%\item [(i)] $\mathbb{P}\{\hat\delta_{q}^*\geq\hat\delta_{q}/\sigma^2\}\rightarrow 1 $;
%\item [(ii)] $\mathbb{P}\{\hat\delta_{i}^*\leq\hat\delta_{i}/\sigma^2\}\rightarrow 1,\ for\ q+1\leq i\leq p-2$;
%\item [(iii)] ${\hat\delta_{q+1}^*}/{\hat\delta_{q}^*}\leq {\hat\delta_{q+1}}/{\hat\delta_{q}}$.
%\end{itemize}
\begin{remark}
Conditions $(i)$ and $(ii)$ ensure the transformation could pull up the value of $\hat\delta_{q}$ and bring down that of $\hat\delta_{i}$, for $q+1\leq i\leq p-2$. Condition (iii) is critical to ensuring that the ``valley'' could be closer to its limit ``0'' and then be better separated from the ``cliff'' after the transformation.
\end{remark}
The following conditions $(a)-(c)$ ensure that  $f_n:\mathbb{R}\rightarrow \mathbb{R}$ satisfies the above requirements $(i)-(iii)$, letting $f_n'(x)$ be the derivative of $f_n(x)$ with respect to $x$:\\
\indent (a) $f_n\ is\ differentiable,\ and\ f_n'\geq 0\ in\ \mathbb{R}$;\\
\indent (b) $f_n'\ is\ increasing\ and\ nonnegative\ in\ \mathbb{R}$;\\
\indent (c) there exists a sequence $ \kappa_n>0,\ \hat\lambda_{i}/\sigma^2-e=o_p(\kappa_n)\ for\ q+1\leq i\leq L-1$ such that $f_n'(x)=1$ for all $ x\in (e-\kappa_n,e+\kappa_n)$.
%\begin{itemize}
%\item [(a)] $f_n\ is\ differentiable,\ and\ f_n'\geq 0\ in\ \mathbb{R}$;
%\item [(b)] $f_n'\ is\ increasing\ and\ nonnegative\ in\ \mathbb{R}$;
%\item [(c)] there exists a sequence $ \kappa_n>0,\ \hat\lambda_{i}/\sigma^2-e=o_p(\kappa_n)\ for\ q+1\leq i\leq L-1$ such that $f_n'(x)=1$ for all $ x\in (e-\kappa_n,e+\kappa_n)$.
%\end{itemize}

\begin{lemma}\label{4}
Conditions $(a)-(c)$ imply Requirements $(i)-(iii)$ for $\{\hat\delta_{n,i}^*\}$ and $\{\hat\delta_{n,i}\}$ defined as above.
\end{lemma}
\begin{remark}
In Condition $(c)$, $\kappa_n$ can take a wide range  of values, as long as it satisfies the condition that $\hat \lambda_i/\sigma^2-e=o_p(\kappa_n)$ for $q+1\leq i\leq L-1$. Specifically, it can take a constant value, converge to zero or even converge to infinity. Let $f_n'$ take value $1$ in a neighbourhood of the  value $e$ with radius $\kappa_n$, so that all $\hat\lambda_{i}/\sigma^2$, for $q+1\le i\le L-1$, fall into  this neighbourhood.
%due to the assumption that $\hat\lambda_{i}/\sigma^2-e=o_p(\kappa_n)$.
%, i.e., $\tilde c_n=o(\kappa_n)$, for $q+1\leq i\leq L-1$
Thus, the ratios $\hat r_i^{TR}$, for $q+1\le i\le L-2$, remain unaffected by the transformation $f_n$. Besides, the selection of $\kappa_n$ is independent of $c_n$.
\end{remark}
%\begin{remark}
We now give a piecewise quadratic function for this purpose as follows: %In Condition $(c)$, we let $f_n'$ take value $1$ in a neighbourhood of the  value $e$ whose radius equals $\theta_n$. Choose a function $f_n'$ from the class of piecewise linear functions. Then
%Let $\{f_n\}_{n\geq1}$ be a piecewise quadratic function:
%\begin{eqnarray}
%f_n'(x)=
%\begin{cases}0,&x< L_n-\frac{1}{k_1}\\
%k_1(x-L_n)+1, &L_n-\frac{1}{k_1}\leq x<L_n\\
%1,&L_n\leq x<R_n\\
%k_2(x-R_n)+1,&x\geq R_n
%\end{cases},
%\end{eqnarray}\\
%where $k_1$ and $k_2$ are slopes to be determined, $L_n=e-\theta_n$, $R_n=e+\theta_n$. Letting the value $e$ be a fixed point of function $f_n$ such that $f_n(e)=e$, we easily get the expression of $f_n$ through $f_n'$ as
\begin{equation}
f_n(x)=\left\{
\begin{array}{lcl}
L_n-\frac{1}{2k_1},&&x< L_n-\frac{1}{k_1}\\
\frac{1}{2}k_1x^2+(1-k_1L_n)x+\frac{1}{2}k_1L_n^2, &&L_n-\frac{1}{k_1}\leq x<L_n\\
x,&&L_n\leq x<R_n\\
\frac{1}{2}k_2x^2+(1-k_2R_n)x+\frac{1}{2}k_2R_n^2,&&x\geq R_n
\end{array} \right.
\end{equation}
%\begin{equation}
%f_n(x)=\left\{
%\begin{array}{rcl}
%
%\left\{
%\begin{aligned}
%L_n-\frac{1}{2k_1},&x< L_n-\frac{1}{k_1}\\
%\frac{1}{2}k_1x^2+(1-k_1L_n)x+\frac{1}{2}k_1L_n^2, &L_n-\frac{1}{k_1}\leq x<L_n\\
%x,&L_n\leq x<R_n\\
%\frac{1}{2}k_2x^2+(1-k_2R_n)x+\frac{1}{2}k_2R_n^2,&x\geq R_n
%\end{aligned}
%\right ,
%\vspace{-0.8em}
%\end{equation}
where $k_1$ and $k_2$ are slopes of $\{f'_n\}_{n\geq1}$ to be determined, $L_n=e-\kappa_n$, $R_n=e+\kappa_n$.
%Also, Condition $(c)$ requires only that $\kappa_n$ have a lower convergence rate than $\hat\lambda_{i}$, for $q+1\leq i\leq L-1$.
It is clear that when $k_1$ and $k_2$ are $0$, the TVACLE  degenerates to the VACLE.

%\end{remark}

%\subsection{Consistency of the TVACLE Estimator}
The consistency of $\hat{q}_n^{TVACLE}$ is stated in the following theorem.
\begin{theorem}\label{3}
Under the same conditions of Theorem \ref{2}, the estimator $\hat{q}_n^{TVACLE}$ with the above transformation $f_n$ is equal to $q$ with a probability going to 1.
\end{theorem}
\begin{remark}
Although selecting an optimal transformation is desirable, we suspect the existence as there are a large class of functions that could satisfy the conditions. Thus, such an issue is  beyond the scope of this paper.
%We will consider whether a criterion for optimal selection or a data-driven selection could be derived in a future study.
%Also, when $\sigma^2$ is unknown, an estimation method is needed. See further discussions in Section~\ref{sec5}. As long as the estimation is consistent at a convergence rate fast enough, the above result is still available.
\end{remark}
%\begin{remark} The main purpose of introducing requirements $(i)-(iii)$ for $\hat\delta_{n,i}^*$ is to alleviate the underestimation difficulty and also to alleviate the sensitivity to the ridge $c_n$. These three conditions are sufficient  to ensure the consistency of $\hat{q}^{TVACLE}$, which can be easily observed in the proof in Appendix. Hence,  conditions $(a)-(c)$ for the transformation functions $\{f_n\}$ are stronger than necessary for the consistency. How to get necessary conditions is still an issue. When $\sigma^2$ is unknown, we need to use an estimator $\hat \sigma^2$ to replace  it. In this paper, we use the one-step estimation procedure in (\ref{13}) for this purpose.
%\end{remark}

\section{Applications to several models}\label{sec4}
In this section, we introduce several special models to which our method can be applied.\\ 

%As the results in Theorem 3.1 are very general under Assumption~2.1. Thus, we in this section verify it for three models.

%\textbf{Factor Model}\quad We consider the following factor model in pervasive multivariate setting, i.e. with fixed $p$.
%\begin{equation}
%Y=AF+\varepsilon,
%\end{equation}
%where $F$ is a $q$-dimensional common factor vector, $A$ is the factor loading matrix of size $p\times q$ and $\varepsilon$ is a $p$-dimensional noise vector with $E(\varepsilon)=0$ and $Var(\varepsilon)=I_p$. Assuming that the common factor is independent of noise, then the covariance matrix of $Y$ is $\Sigma_Y=AVar(F)A'+I_p:=A\Sigma_FA'+I_p$ can be regarded as a $q$-dimensional perturbation of the identity matrix $I_p$, and $A\Sigma_FA'$ is the perturbation matrix with rank $q$. Generally, the $q\times q$ symmetric matrix $A\Sigma_FA'$ is of full rank $q$, so it has $q$ positive eigenvalues, and then those $q$ largest eigenvalues of the matrix $\Sigma_Y$ are exactly separated from the rest $p-q$ unit eigenvalues. Let $\hat{\Sigma}_Y$ be the sample counterpart of $\Sigma_Y$, then all these eigenvalues $\{\lambda_{n,i}\}_{1\leq i\leq p}$ of $\hat{\Sigma}_Y$ converge to their counterparts $\{\alpha_i\}_{1\leq i\leq p}$ of $\Sigma_Y$, which is implied by the law of large number. In other words, $\lambda_{n,i}$ converges to $1$ for $q+1\leq i\leq p$ but $\lambda_{n,q}$ converges to $\alpha_q>1$, Thus, assumption \ref{1} holds for this case,  and then TVACLE criterion can be used to determine the number of factors $q$.\\

\textbf{Spiked population models}\quad The model can also be motivated from  the signal detection problem (see, e.g. \cite{nadler2010nonparametric}):
\begin{equation}\label{12}
x_i=\mathbf{A}u_i+\varepsilon_i,\ 1\le i\le n,
\end{equation}
where $u_i\in \mathbb{R}^q$ is a $q$-dimensional random signal vector with zero mean components, $\varepsilon_i\in \mathbb{R}^p$ is a $p$-dimensional random vector with mean zero and covariance matrix $\sigma^2\mathbf{I}_p$
%where $\mathbf{I}_p$ is a $p$-dimensional identity matrix
, $\mathbf{A}\in \mathbb{R}^{p\times q}$ is the steering matrix whose $q$ columns are linearly independent of each other and $x_i\in \mathbb{R}^p$ is the observed vector on the $p$ sensors. Assume that the covariance matrix of the signals is of full rank with $q$ largest eigenvalues $\lambda_1\ge \cdots\ge \lambda_q>\sigma^2>0$. With the typical high dimensional setting in which $p/n\rightarrow c\in (0,+\infty)$, the population covariance matrix $\Sigma_p$ of $x_i$ coincides with the structure of a spiked population model:
\begin{equation}
\text{spec}(\Sigma_p)=\{\lambda_1,\cdots,\lambda_q,\sigma^2,\cdots,\sigma^2\}, 
\end{equation}
Theoretically, the spiked population model allows the existence of small spikes (i.e. $\lambda_i<\sigma^2$), but this case is not discussed in this paper.
\begin{theorem}\label{4.1}
Suppose that $\lambda_q>U(F)=\sigma^2(1+\sqrt c)$. This bound is optimal for the identifiability of $q$. Model Feature~\ref{1} then holds, and the results of Theorems~\ref{2} and \ref{3} hold true.
\end{theorem}
\begin{remark}
From the proof in the supplementary materials, we see the optimality of the lower bound $U(F)=\sigma^2(1+\sqrt c)$ because it is not possible to identify any eigenvalue $\lambda_i$ such that $\sigma^2(1+\sqrt c) >\lambda_i>\sigma^2$ for any $q< i \le L$.
\end{remark}

%Since the spiked population model is a typical and concise model in high-dimensional paradigms, we  consider it to examine the performance of TVACLE and compare it with the estimators defined by Passemier and Yao (2012, 2014) in the simulation studies later.  More discussions and solutions about this problem can be founded in, for example, Kritchman and Nadler (2008), Ulfarsson and Solo (2008), Nadler (2010), Passemier and Yao (2012, 2014).

\vspace{0.4cm}
\textbf{Large-dimensional spiked Fisher matrix}\quad  Again consider the signal detection problem discussed above,
\begin{equation}
x_i=\mathbf{A}u_i+\varepsilon_i,\ 1\le i\le n,
\end{equation}
where $x_i$, $\mathbf{A}$ and $u_i$ share the same settings of (\ref{12}), whilst $\varepsilon_i$ is  a noise vector with a general covariance matrix $\Sigma_2$. Denote the population covariance matrix of $x_i$ by $\Sigma_1$ such that
$ \Sigma_1=\Sigma_2+\Delta,$
where $\Delta=\mathbf{A}cov(u_i)\mathbf{A}^T$ is a non-negative definite matrix with fixed rank $q$ provided that $cov(u_i)$ is of full rank. If $\Sigma_1=\sigma^2\mathbf{I}_p$, it degenerates to the spiked population model.
%Consider two $p$-dimensional populations with  covariance matrices $\Sigma_1$ and $\Sigma_2$ respectively with
Otherwise, we note that $\Sigma_1\Sigma_2^{-1}$ has a spiked structure as
\begin{equation}
\text{spec}(\Sigma_1\Sigma_2^{-1})=\{\lambda_1,\cdots,\lambda_{q},1,\cdots,1\},
\end{equation}
where $\lambda_1\geq\cdots\ge\lambda_{q}>1$ and the number of spikes $q$ is fixed.
Let $\mathbf{S}_1$ and $\mathbf{S}_2$ %$\hat\Sigma_1$ and $\hat\Sigma_2$
be the sample covariance matrices that correspond to $\Sigma_1$ and $\Sigma_2$ with respective sample sizes of $n$ and $T$, where $p/n\rightarrow c>0$ and $p/T\rightarrow y\in (0,1)$ as $n\to \infty$ and $T\to \infty$ respectively. %For the hypothesis test
%\begin{eqnarray}
%H_0:\ \Sigma_1=\Sigma_2\quad
%H_1:\ \Sigma_1=\Sigma_2+\Delta,
%\end{eqnarray}
%where $\Delta$ is a perturbation matrix with fixed rank $q$. %$\hat\Sigma_2$
Note that there are two different sample sizes $n$ and $T$ because the sample covariance matrix $\mathbf{S}_2$ comes from another sequence of pure noise observations, say $\{e_i\}_{1\le i\le T}$, with a different sample size $T$. Denote $\mathbf{S}_1=\frac{1}{n}\sum_{i=1}^nx_ix_i^{\top}$ and $\mathbf{S}_2=\frac{1}{T}\sum_{i=1}^ne_ie_i^{\top}$. When $\mathbf{S}_2$ is invertible, the random matrix $\mathbf{F}_n=\mathbf{S}_1 \mathbf{S}_2^{-1}$ is called a Fisher matrix,
whose motivation comes from the following hypothesis testing problem:
\begin{equation}
H_0:\ \Sigma_1=\Sigma_2\quad
H_1:\ \Sigma_1=\Sigma_2+\Delta.
\end{equation}
%Consider a Fisher matrix with spiked structure of the form
%\begin{equation}
%\text{spec}(\Sigma_1\Sigma_2^{-1})=\{\lambda_1,\lambda_2,\cdots,\lambda_{q_1},1,\cdots,1\},
%\end{equation}
%where $\lambda_1\geq\lambda_2\geq\cdots\ge\lambda_{q_1}>1$ and the number of spikes $q$ is fixed.
See \cite{wang2017extreme} as an example. Denote the eigenvalues of $\mathbf{F}_n$ as $\hat\lambda_{1}\geq\cdots\ge\hat\lambda_{p}$. The difference between the two hypotheses then relies upon those extreme eigenvalues of $\mathbf{F}_n$.

We consider a more general Fisher matrix with the spiked structure
\begin{equation}
\text{spec}(\Sigma_1\Sigma_2^{-1})=\{\lambda_1,\cdots,\lambda_{q},\sigma^2,\cdots,\sigma^2\}. 
\end{equation}
This is motivated by the hypothesis testing problem:
\begin{equation}
H_0:\ \Sigma_1=\sigma^2\Sigma_2\quad
H_1:\ \Sigma_1=\sigma^2\Sigma_2+\Delta, 
\end{equation}
Also, by using the simple transformation $\hat\lambda_{i}\rightarrowtail (\sigma^2)^{-1}\hat\lambda_{i}$, we can achieve the results in the case of $\sigma^2=1$ in a similar manner.
\begin{theorem}\label{4.3}
Suppose that $\lambda_q>U(F)=\sigma^2\gamma(1+\sqrt{c+y-cy})$, where $\gamma=(1-y)^{-1}$. This is the optimal lower bound for identifiability of $q$, and Model Feature~\ref{1} holds. The results of Theorems~\ref{2} and \ref{3} hold true.
\end{theorem}

In the following section, we consider the auto-covariance matrix. This matrix has a more much complicated structure at the sample level, so we provide some more discussion that does not exactly follow the examination for the Model Feature~~\ref{1} in Theorems~\ref{2} and \ref{3}. In addition, because the theoretical analysis for the estimated matrix is not as complete as those for the spiked population and Fisher matrix, we must then add some extra assumptions on the convergence rate of the estimated eigenvalues if we wish to derive the estimation consistency, although it would be true, as reasonably conjectured by \cite{li2017identifying}. This requires a rigorous proof, but it is beyond the scope of this paper. We shall therefore leave it to a further study. In this paper, however, we provide a proposition that assumes that the convergence rate can be achieved and use numerical studies to verify the usefulness of our method in practice.\\

\textbf{Large dimensional auto-covariance matrix}\quad Consider a factor model: % for high-dimensional time series in \cite{lam2012factor},
\begin{equation}\label{auto-model}
y_t=\mathbf{A}x_t+\varepsilon_t,
\end{equation}
where for a fixed number $q_0$, $x_t$ is a $q_0$-dimensional common factor time series, $\mathbf{A}$ is the $p\times q_0$ factor loading matrix, $\{\varepsilon_t\}$ is a sequence of Gaussian noise independent of $x_t$ and $y_t$ is the $t$-th column of the $p\times T$ observed matrix $\mathbf{Y}$; namely, $p$-dimensional observation at time $t$.
Let $\Sigma_y=cov(y_t,y_{t-1})$ be the lag-1 auto-covariance matrices of $y_t$, and let $\hat{\Sigma}_y=\frac{1}{T}\sum_{t=2}^{T+1}y_ty_{t-1}^{\top}$ be its sample version.
%Since  $\hat{\Sigma}_y$ is an asymmetric matrix, \cite{li2017identifying} considered its singular values. This is equivalent to considering the square root of the eigenvalues of the matrices $\hat{M}_y:=\hat{\Sigma}_y\hat{\Sigma}_y'$.

Let $\mu$ be a finite measure on the real line $\mathbb{R}$ with support denoted by $supp(\mu)$ and $\mathbb{C}\backslash supp(\mu)$ be a complex space $\mathbb{C}$ subtracting the set $supp(\mu)$. For any $z \in \mathbb{C}\backslash supp(\mu)$, the Stieltjes transformation and T-transformation of $\mu$ are respectively defined as
\begin{equation}\label{trans}
\mathcal{S}(z)=\int\frac{1}{t-z}d\mu(t),\ \mathcal{T}(z)=\int\frac{t}{z-t}d\mu(t).
\end{equation}
When $\mu$ is supported on an interval, say $supp(\mu)=[A,B]$, and $z$ is a real value, the $T$-transformation $\mathcal{T}(\cdot)$ is a decreasing homeomorphism from $(-\infty, A)$ onto $(\mathcal{T}(A-), 0)$ and from $(B, +\infty)$ onto $(0, \mathcal{T}(B+))$, where
$$\mathcal{T}(A-):=\lim_{z\in \mathbb{R}, z\rightarrow A-}\mathcal{T}(z), \
\mathcal{T}(B+):=\lim_{z\in \mathbb{R}, z\rightarrow B+}\mathcal{T}(z).$$

Give the assumptions on the time series $\{x_t\}_{1\le t\le T}$ and $\{\varepsilon_t\}_{1\le t\le T}$ (\cite{li2017identifying}) as follows.
\begin{assumption}\label{auto-assumption1}
$\{x_t\}_{1\le t\le T}$ is a $q_0$-dimensional stationary time series, where $q_0$ is a fixed number. Every component is independent of the others,
$$x_{i,t}=\sum_{l=0}^\infty\alpha_{i,l}\eta_{i,t-l},\ i=1,\cdots,q_0,\ t=1,\cdots,T,$$
where $\{\eta_{i,k}\}$ is a real-valued and weakly stationary white noise with mean $0$ and variance $\sigma_i^2$. Denote $\gamma_0(i)$ and $\gamma_1(i)$ as the variance and lag-1 auto-covariance of $\{x_{i,t}\}$, respectively.
\end{assumption}
\begin{assumption}\label{auto-assumption2}
$\{\varepsilon_t\}$ is a $p$-dimensional real-valued random vector independent of $\{x_t\}$ and with independent components $\varepsilon_{i,t}$, satisfying $\mathbb{E}(\varepsilon_{i,t})=0,\ \mathbb{E}(\varepsilon_{i,t}^2)=\sigma^2,$
and $\forall \eta>0$,
$$\frac{1}{\eta^4p^T}\sum_{i=1}^p\sum_{t=1}^{T+1}\mathbb{E}(|\varepsilon_{i,t}|^4I_{(|\varepsilon_{i,t}\ge\eta T^{1/4}|)})\longrightarrow 0\quad as\ pT\rightarrow \infty.$$
\end{assumption}
In the high dimensional setting with $p/T\rightarrow y>0$, the following result holds.
\begin{prop}\label{4.2}
Denote $\mathcal{T}(\cdot)$ as the $T$-transformation of the limiting spectral distribution for matrix $\hat {\mathbf{M}}_y/\sigma^4\equiv \hat{\Sigma}_y\hat{\Sigma}_y^{\top}/\sigma^4$. Suppose that the above assumptions are satisfied. Let $q=\#\{i: 1\le i\le q_0, \ \mathcal{T}_1(i)<\mathcal{T}(b_1+)\}$, where
$$\mathcal{T}_1(i)=\frac{2y\sigma^2\gamma_0(i)+\gamma_1(i)^2-\sqrt{(2y\sigma^2\gamma_0(i)+\gamma_1(i)^2)^2-4y^2\sigma^4(\gamma_0(i)^2-\gamma_1(i))^2}}{2\gamma_0(i)^2-2\gamma_1(i)^2},$$
$$b_1=(-1+20y+8y^2+(1+8y)^{3/2})/8,\ \mathcal{T}(b_1+)=\lim_{z\in \mathbb{R}, z\rightarrow b_1+}\mathcal{T}(z).$$
Then $q$ is the largest number of common factors that are identifiable. %The results of Theorems~2.1 and 3.1 hold true.
\end{prop}
\begin{remark} Although the constraint $\ \mathcal{T}_1(i)<\mathcal{T}(b_1+)$ does not have a simple formulation presented in Model Feature~\ref{1}, it also provides the optimal bound.
\end{remark}
\begin{prop}\label{4.31}
If the estimated eigenvalues $\hat \lambda_i$ for $i>q$ have a convergence rate of order $O_p(n^{-2/3})$ with the assumptions in Proposition~\ref{4.2}, Model Feature~\ref{1} then holds, and Theorems~\ref{2} and \ref{3} hold true.
\end{prop}
\begin{remark} As we commented before, \cite{li2017identifying} considered a criterion with a reasonable conjecture on the convergence rate of order $O_p(n^{-2/3})$, although without a rigorous proof. We have not provided this result either, and thus we regard the above result as a proposition, rather than a theorem. We will see that it works well in the numerical studies.
\end{remark}
\section{Numerical Studies}\label{sec5}

\subsection{ Scale estimation}\label{sec5.1}
\cite{passemier2012determining} estimated $\sigma^2$ by simply taking the average over $\{\hat{\lambda}_{i}\}_{q+1\le i\le p}$ and \cite{ passemier2017estimation} established its consistency and further introduced a refined version by subtracting the bias. That, however, involves an iteration procedure because the number $q$ must be estimated. To construct a robust estimator, \cite{ulfarsson2008dimension} and \cite{johnstone2009consistency} used the median of the sample eigenvalues $\{\hat{\lambda}_{i}:\hat{\lambda}_{i}\le b\}$ and the sample variances $\{\frac{1}{n}\sum_{i=1}^nx_{ij}^2\}_{1\le j\le p}$, respectively. The former median still requires a crude estimator of the right edge $b=\sigma^2(1+\sqrt c)^2$ in advance, which is equivalent to give a rough initial estimator of $\sigma^2$.

%Notice that $\hat{r}_{n,i}^R$'s dependence on  $\sigma^2$ is only through the selection of ridge $c_n$, which actually does not require a very accurate estimator.
In this section, we propose  %simplify the estimation procedure in Ulfarsson and Solo (2008) without estimating $b$ to have a
a one-step procedure that could be regarded as a simplified version of the method of \cite{ulfarsson2008dimension}. With spiked population models, the empirical spectral distribution of $\mathbf{S}_n$ almost surely converges to a Marcenko-Pastur distribution $F_{c,\sigma^2}(x)$ (see details in Supplementary Materials). For $0<\alpha<1$, their $\alpha$-quantiles are denoted $\hat\xi_{c,\sigma^2}^{(n)}(\alpha)$ and $\xi_{c,\sigma^2}(\alpha)$, respectively:
\begin{equation}
\hat\xi_{c,\sigma^2}^{(n)}(\alpha):=\hat{\lambda}_{p-[p\alpha]},\quad
\xi_{c,\sigma^2}(\alpha):=\inf\{x:F_{c,\sigma^2}(x)\ge\alpha\}.
\end{equation}
It then follows that $\hat\xi_{c,\sigma^2}^{(n)}(\alpha)\rightarrow\xi_{c,\sigma^2}(\alpha),\ as\  n\rightarrow\infty$.
Note that
$\xi_{c,\sigma^2}(\alpha)=\sigma^2\xi_{c,1}(\alpha)$. Approximating a certain quantile, say $\xi_{c,\sigma^2}(\alpha)$, of the M-P distribution by its sample counterpart  $\hat\xi_{c,\sigma^2}^{(n)}(\alpha)$, we obtain an  estimator of $\sigma^2$,
\begin{equation}\label{13}
\hat{\sigma}^2=\hat\xi_{c,\sigma^2}^{(n)}(\alpha)\cdot{\xi_{c,1}(\alpha)}^{-1}.
\end{equation}
% because $\xi_{c,1}(\alpha)$ is a given value.
The consistency of $\hat \sigma^2$ is equivalent to that of $\hat\xi_{c,\sigma^2}^{(n)}(\alpha)$, which can hold under certain conditions.
Further, the rigidity of the eigenvalues of covariance matrix (see Theorem~3.3 in \cite{pillai2014universality}) implies that the convergence rate of $\hat \sigma^2$ is $o(n^{-1+\varepsilon})$ for any $\varepsilon>0$. Thus, consistencies still hold for VACLE and TVACLE when we replace $\hat\lambda_i/\sigma^2$ by $\hat\lambda_i/\hat \sigma^2$, as $\hat\sigma^2$ possesses a higher convergence rate than those extreme eigenvalues $\hat\lambda_{i},\ 1\le i\le L$ for any fixed $L$. %Specifically, $\hat \lambda_i/\sigma^2$ and $\hat \lambda_i/\hat\sigma^2,\ 1\le i\le L$, share the same convergence rate to their limits, and then convergence rates of $\hat\delta_i$ and $\hat\delta_i^*$
Practically, for the sake of simplicity and stability, we let $\alpha=0.5$ for $0<c<1$; and $1-(2c)^{-1}$ for $c\ge 1$. Then $\alpha=1-(2\max\{1,c\})^{-1}$. The sample quantile $\hat\xi_{c,\sigma^2}^{(n)}(\alpha)$ divides all positive eigenvalues of $\mathbf{S}_n$ into two equal parts. The estimator $\hat{\sigma}^2$  is then less sensitive to extreme eigenvalues of $\mathbf{S}_n$. Its performance is examined in the following numerical studies.

\subsection{ Simulations about spiked population models}\label{sec5.2}
In this subsection, we consider the comparisons between VACLE and TVACLE defined as $\hat q_n^{VACLE}$ and $\hat q_n^{TVACLE}$ and the method defined as $\hat q_n^{PY}$ developed and refined by \cite{passemier2012determining} and \cite{passemier2014estimation}. Because estimating the number of spikes $q$ is the main focus, we conduct the simulations mainly with given $\sigma^2$. For unknown $\sigma^2$, we give a brief discussion, and a simple one-step estimator of $\sigma^2$ is introduced in Section~\ref{sec5.1}. In all simulation experiments, we conduct $500$ independent replications. %Note that the TVACLE attempts to alleviate the sensitivity of ridge selection, but the construction is more complicated than that of the VACLE. We then also include a comparison between these two criteria to determine their performances. % such that a good recommendation can be made.
Furthermore, recalling the definition of $c=p/n$, we report the results with three scenarios: $c=.25, 1$ and $2$ to represent the cases with dimensions $p$ smaller than and larger than the sample size $n$, respectively. %We also conducted simulations with more scenarios that are not reported in this paper because the conclusions were very similar.
%\subsubsection{A Brief Review on Passemier and Yao's Method }
%In this part, we recall the estimation criteria in Passemier and Yao (2012 and 2014). They share the same notations and assumptions with above sections.
$\hat q_n^{PY}$ is defined by % as it is an improved version of  Passemier and Yao (2012), as we mentioned in the introduction,
%which is a scree plot-based criterion:
\begin{equation}\label{}
\hat{q}_n^{PY}=min\{i\in \{1,\cdots ,Q\}:\hat\delta_{i+1}<d_n \ and\ \hat\delta_{i+2}<d_n\}, 
\end{equation}
where $L>q$ is a prefixed bound large enough, $d_n=o(n^{-1/2})$ and $n^{2/3}d_n\rightarrow+\infty$. %Under some regularity conditions, they checked the consistency of $\hat{q}_n^{PY}$.

%\begin{remark}
%Note that at the sample level, the differences  $\hat\delta_{n,i}$ derived from nonspikes and those from equal spikes could be very close to each other in values. Thus,  the threshold $d_n$ could not separate them well, especially in the case  with high, even moderate multiplicity. Then  $\hat{q}_n^{PY}$ could be smaller than the actual one because when  two consecutive $\hat\delta_{n,i}$ from equal spikes are both below the threshold $d_n$, this happens. This underestimation problem is exposed in our simulation experiments below.
%\end{remark}

\subsubsection*{Models and parameters selections: the known $\sigma^2$ case.}
%First, we select a sequence of threshold values.
For $\hat{q}_n^{PY}$, the sequence $d_n=Cn^{-2/3}\sqrt{2loglogn}$ with $C$ being  adjusted by an automatic procedure identical to that in \cite{passemier2014estimation}.  For $\hat{q}_n^{VACLE}$ and $\hat{q}_n^{TVACLE}$, they share the same threshold $\tau=0.5$ but have different ridges $c_n$. Theoretically speaking, the selection range of $c_n$ is very wide to meet the requirement that  $c_n\rightarrow 0$ and $n^{2/3}c_n\rightarrow +\infty$. Here, we give an automatic procedure for ridge calibration by pure-noise simulations.
For given $(p,n)$, we conduct 500 independent pure-noise simulations and obtain the $\alpha-$quantile $\text{q}_{p,n}(\alpha)$ and sample mean $\text{m}_{p,n}$ of the difference  $\{\hat \lambda_1- \hat \lambda_2\}$, where $\hat \lambda_1$ and $\hat \lambda_2$ are the two largest eigenvalues of the noise matrix. By results in \cite{benaych2011fluctuations}, for such a no-spike matrix, we can use $\{\hat \lambda_1-\hat \lambda_2\}$ to approximate $\hat \delta_{q+1}$:
\begin{eqnarray*}
&&\mathbb{P}\{\text{q}_{p,n}(0.01)-m_{p,n}<\hat\delta_{q+1}-m_{p,n}<\text{q}_{p,n}(0.99)-m_{p,n}\}\nonumber\\
&\approx& \mathbb{P}\{\text{q}_{p,n}(0.01)-m_{p,n}<\hat \lambda_1-\hat \lambda_2-m_{p,n}<\text{q}_{p,n}(0.99)-m_{p,n}\}\approx 0.98.
\end{eqnarray*}
Thus, the ratio $\left[\hat\delta_{q+2}-m_{p,n}+[\text{q}_{p,n}(0.99)-\text{q}_{p,n}(0.01)]\right]\left[\hat\delta_{q+1}-m_{p,n}+[\text{q}_{p,n}(0.99)-\text{q}_{p,n}(0.01)]\right]^{-1}$ would be dominated by the term $[\text{q}_{p,n}(0.99)-\text{q}_{p,n}(0.01)-m_{p,n}]$ and then remain close to the ``cliff'' valued at $1$ with a high probability. To ensure the convergence rate, we select ridge $c_n^{(1)}=\log\log n\cdot[\text{q}_{p,n}(0.95)-\text{q}_{p,n}(0.05)]-\text{m}_{p,n}$ for  the VACLE  and a smaller one $c_n^{(2)}=\sqrt{\log\log n}[\text{q}_{p,n}(0.95)-\text{q}_{p,n}(0.05)]-\text{m}_{p,n}$ for the TVACLE. Note that $\text{q}_{p,n}(\alpha)$ and $\text{m}_{p,n}$ converge to zero at the same rate as $\hat\lambda_{q+1}$ that has a slightly faster rate to zero than $c_n^{(1)}$ and $c_n^{(2)}$.
%By the rule of thumb, we select the slowly convergent sequence $D\cdot (log_2n)^{-1}$, where $D$ is a constant taking value 2 and 1 in $\hat{q}_n^{VACLE}$ and $\hat{q}_n^{TVACLE}$, respectively.
In addition, we manually determine the sequence $\kappa_n$. Details in the parameters selections are reported in Table~\ref{spiked parameter1}. %in Table~\ref{table1}.
Following the calibration procedure of \cite{passemier2014estimation}, we obtain the value of $C$ for various $c=p/n$, as shown in Table~\ref{spiked parameter2}.
%From the calibration procedure of \cite{passemier2014estimation}, the tuning parameter $C$ is reported in the table above. %~\ref{table1}. Following their process, we obtain the value of $C$ for various $c=p/n$, as shown in the following Table~\ref{spiked parameter2}.%~\ref{table2}:
\begin{table}[H]\footnotesize
\caption{\footnotesize Parameters settings for the three methods.}
\centering
\renewcommand\arraystretch{0.8}
\begin{tabular}{cccccccc}
\toprule
Method & $d_n$ &$\tau$& $c_n$ & $\kappa_n$ &$k_1$ &$k_2$ &$L$\\
\midrule
PY & $C\cdot n^{-2/3}\sqrt{2\log\log n}$ & | &|&|&|&|&20\\
VACLE & | & 0.5 &$c_n^{(1)}$&|&|&|&20\\
TVACLE & | & 0.5 &$c_n^{(2)}$&$\log\log p \cdot p^{-2/3}$&5&5&20\\
\bottomrule
\end{tabular}
\label{spiked parameter1}
\end{table}
\begin{table}[H]\footnotesize
\caption{\footnotesize Values of $C$ .}
\centering
\renewcommand\arraystretch{0.8}
\begin{tabular}{p{25pt}p{40pt}p{40pt}p{40pt}p{40pt}}
\toprule
{\it c=p/n} & 0.25 %&0.5
&1&2 \\
\midrule
C &5.5226 %&5.8953
&6.3424&7.6257\\
\bottomrule
\end{tabular}
\label{spiked parameter2}
\end{table}
\begin{remark}
Note that we select different ridges $c_n$ in $\hat{q}_n^{VACLE}$ and $\hat{q}_n^{TVACLE}$. As described above, a relatively small ridge $c_n$ could cause $\hat r_{q+1}^R$ to have better separation from $\hat r_{q}^R$, but it might also result in instability of $\hat r_{i}^R$ for $i>q+1$, so a relatively large ridge is needed for $\hat r_{i}^R$. However, $\hat r_{i}^{TR}$ would be much less sensitive to the ridge. Thus, we choose a smaller ridge for $\hat{q}_n^{TVACLE}$. Besides, ridges $c_n^{(1)}$ and $c_n^{(2)}$ are generated by an automatic procedure instead of manual selections. This calibration procedure only depends on $(p,n)$. % and it is completely independent of settings of those spikes.
\end{remark}
Consider three models: for fair comparison, Models~\ref{model1} and \ref{model2} were used by \cite{passemier2012determining} with dispersed spikes and closely spaced but unequal spikes respectively, and Model~\ref{model3} has two equal spikes:
\begin{model}\label{model1}
$q=5$, $\left(\lambda_1,\cdots,\lambda_5\right)=(259.72,17.97,11.04,7.88,4.82)$,
\end{model}
\begin{model}\label{model2}
$q=4$, $\left(\lambda_1,\cdots,\lambda_4\right)=(7,6,5,4)$,
\end{model}
\begin{model}\label{model3}
$q=4$, $\left(\lambda_1,\cdots,\lambda_4\right)=(5,4,3,3)$.
\end{model}
%\textbf{Model 1}: $q=5$, $\left(\lambda_1,\cdots,\lambda_5\right)=(259.72,17.97,11.04,7.88,4.82)$,\\
%\indent \textbf{Model 2}: $q=4$, $\left(\lambda_1,\cdots,\lambda_4\right)=(7,6,5,4)$,\\
%\indent \textbf{Model 3}: $q=4$, $\left(\lambda_1,\cdots,\lambda_4\right)=(5,4,3,3)$.\\
Furthermore, we compare $\hat{q}_n^{TVACLE}$ with $\hat{q}_n^{PY}$ on a model  with greater multiplicity of spikes:
\begin{model}\label{model4}
$q=6$, $\left(\lambda_1,\cdots,\lambda_6\right)=(5,5,5,5,5,5)$.
\end{model}
%\indent \textbf{Model 4}: $q=6$, $\left(\lambda_1,\cdots,\lambda_6\right)=(5,5,5,5,5,5)$.\\
Set $\sigma^2=1$. When $\sigma^2$ is regarded as unknown, use the  one-step method in $(\ref{13})$ to estimate it. We conduct the same simulations for $\hat{q}_n^{VACLE}$ and $\hat{q}_n^{TVACLE}$ as those with the known $\sigma^2$, but we do not report the results of $\hat{q}_n^{PY}$ with the unknown $\sigma^2$ because we found that the results and conclusions are very similar.

\subsubsection*{Numerical Performance for Known $\sigma^2$ Case}
\begin{table}[H]\footnotesize
\caption{\footnotesize
Mean, mean square error and misestimation rates of $\hat{q}_n^{PY}$, $\hat{q}_n^{VACLE}$ and $\hat{q}_n^{TVACLE}$ over 500 independent replications for Models~\ref{model1}-\ref{model3}, with known $\sigma^2=1$.}
\centering
{\small\scriptsize\hspace{12.5cm}
\renewcommand{\arraystretch}{1}\tabcolsep 0.1cm
\renewcommand\arraystretch{0.8}
\begin{tabular}{c|c|ccc|ccc|ccc}
\hline
& & \multicolumn{3}{c}{$\hat{q}_n^{PY}$} & \multicolumn{3}{c}{$\hat{q}_n^{VACLE}$} &\multicolumn{3}{c}{$\hat{q}_n^{TVACLE}$}  \\
& $(p,n)$& \multicolumn{1}{c}{Mean}&\multicolumn{1}{c}{MSE}&\multicolumn{1}{c}{$\hat{q}_n^{PY}\neq q$}&\multicolumn{1}{c}{Mean}&\multicolumn{1}{c}{MSE}&\multicolumn{1}{c}{$\hat{q}_n^{VACLE}\neq q$}&\multicolumn{1}{c}{Mean}&\multicolumn{1}{c}{MSE} &\multicolumn{1}{c}{$\hat{q}_n^{TVACLE}\neq q$} \\
\hline
\multirow{6}{*}{Model~\ref{model1}}
&$(50,200)$  &5.022  &0.022   &0.022   &5.004   &0.004  &0.004  &5.024  &0.024 &0.024\\
&$(200,800)$  &5.012 &0.012  &0.012  &5.002 &0.002  &0.002 &5.016  &0.016 &0.016 \\
\cline{2-11}
%$(50,100)$  &5.02  &0.024 &0.018 &5.026  &0.034   &0.022  &5.016   &0.016 &0.016\\
%$(200,400)$  &5.022  &0.022   &0.022 &5  &0   &0  &5.036   &0.036 &0.036\\
%\hline
&$(100,100)$ &5.016  &0.02   &0.02   &4.97   &0.046    &0.046   &4.998    &0.002   &0.002 \\
&$(200,200)$  &5.026  &0.03   &0.024   &5.01   &0.01 &0.01  &5.004 &0.004 &0.004 \\
\cline{2-11}
&$(100,50)$ &4.846  &0.218 &0.212 &4.484   &1.296 &0.41 &4.782 &0.222 &0.216 \\
&$(200,100)$ &4.99  &0.074   &0.074   &4.758 &0.486 &0.194 &4.954  &0.046 &0.046 \\
\hline
\multirow{6}{*}{Model~\ref{model2}}
&$(50,200)$ &4.018 &0.058 &0.028 &4.006 &0.006 &0.006 &4.016 &0.016 &0.016 \\
&$(200,800)$ &4.016 &0.02 &0.014 &4.004 &0.004 &0.004 &4.032 &0.04 &0.028 \\
\cline{2-11}
%$(50,100)$ &4.006  &0.054   &0.022 &3.948  &0.284   &0.06  &4.014  &0.014 &0.014\\
%$(200,400)$ &4.018  &0.018   &0.018 &4  &0   &0  &4.044   &0.044 &0.044\\
%\hline
&$(100,100)$ &3.922 &0.246 &0.074 &3.416 &2.112 &0.22 &3.968 &0.036 &0.036 \\
&$(200,200)$ &4.014 &0.014 &0.014 &3.92 &0.304 &0.048 &4.006 &0.006 &0.006 \\
\cline{2-11}
&$(200,100)$ &3.558 &0.83 &0.342 &2.452 &5.144 &0.584 &3.712 &0.304 &0.28 \\
&$(400,200)$ &3.906 &0.162 &0.118 &3.046 &3.138 &0.364 &3.958 &0.05 &0.044 \\
\hline
\multirow{6}{*}{Model~\ref{model3}}
&$(50,200)$ &3.994 &0.118 &0.032 &3.772 &0.804 &0.08 &4.024 &0.024 &0.024 \\
&$(200,800)$ &4.018 &0.018 &0.018 &4 &0 &0 &4.036 &0.036 &0.036 \\
%\hline
%$(50,100)$ &3.52  &1.016   &0.33 &2.522  &5.438  &0.556  &3.902   &0.126 &0.114\\
%$(200,400)$ &4.008  &0.008   &0.008 &3.746  &0.886  &0.08  &4.028   &0.028 &0.028\\
\cline{2-11}
&$(200,200)$ &3.456 &0.92 &0.414 &1.94 &6.684 &0.734 &3.614 &0.518 &0.326 \\
&$(400,400)$ &3.904 &0.18 &0.122 &2.7 &4.152 &0.478 &3.898 &0.142 &0.112 \\
\cline{2-11}
&$(400,200)$ &2.222 &3.81 &0.952 &1.08 &9.736 &0.968 &2.648 &2.296 &0.91 \\
&$(800,400)$ &2.626 &2.482 &0.844 &1.588 &7.104 &0.954 &3.022 &1.558 &0.7 \\
\hline
\end{tabular}
}
\label{table1}
\end{table}
From Table~\ref{table1}, we have the following observations. For Model~\ref{model1}, all three methods work well with high accuracies and small MSE in the cases where the dimension $p$ is smaller than $n$ ($c=p/n =0.25$). % and $c=0.5$).
 When either $c=1$ or $c=2$, $\hat{q}_n^{TVACLE}$ is the best, and $\hat{q}_n^{PY}$ also has smaller MSEs than $\hat{q}_n^{VACLE}$. In a word, all three methods perform in a satisfactory manner, but the performance of $\hat{q}_n^{TVACLE}$ is the most stable for various ratio of  $c=p/n$. For Model~\ref{model2}, $\hat{q}_n^{VACLE}$ is sensitive to the ratio $c$, particularly its MSE. Although when $c=2$, $\hat{q}_n^{TVACLE}$ may sometimes slightly underestimate the true number, it is less serious than $\hat{q}_n^{PY}$.
For Model~\ref{model3} with two equal spikes, $\hat{q}_n^{TVACLE}$ works much better than both $\hat{q}_n^{PY}$ and $\hat{q}_n^{VACLE}$ that underestimate $q$ significantly.

\begin{figure}[H]
\centering
\caption{\footnotesize Plots of the first 40 differences and ratios: the left is for differences $\hat\delta_i$, $1\le i\le 40$, in $\hat q_n^{PY}$; the right is for ratios $\hat r_i^{TR}$, $1\le i\le 40$, in $\hat q_n^{TVACLE}$. The results are based on simulations for Model~\ref{model4} with 500 independent replications, and $(p,n)=(400,200)$.}
\includegraphics[scale=0.5]{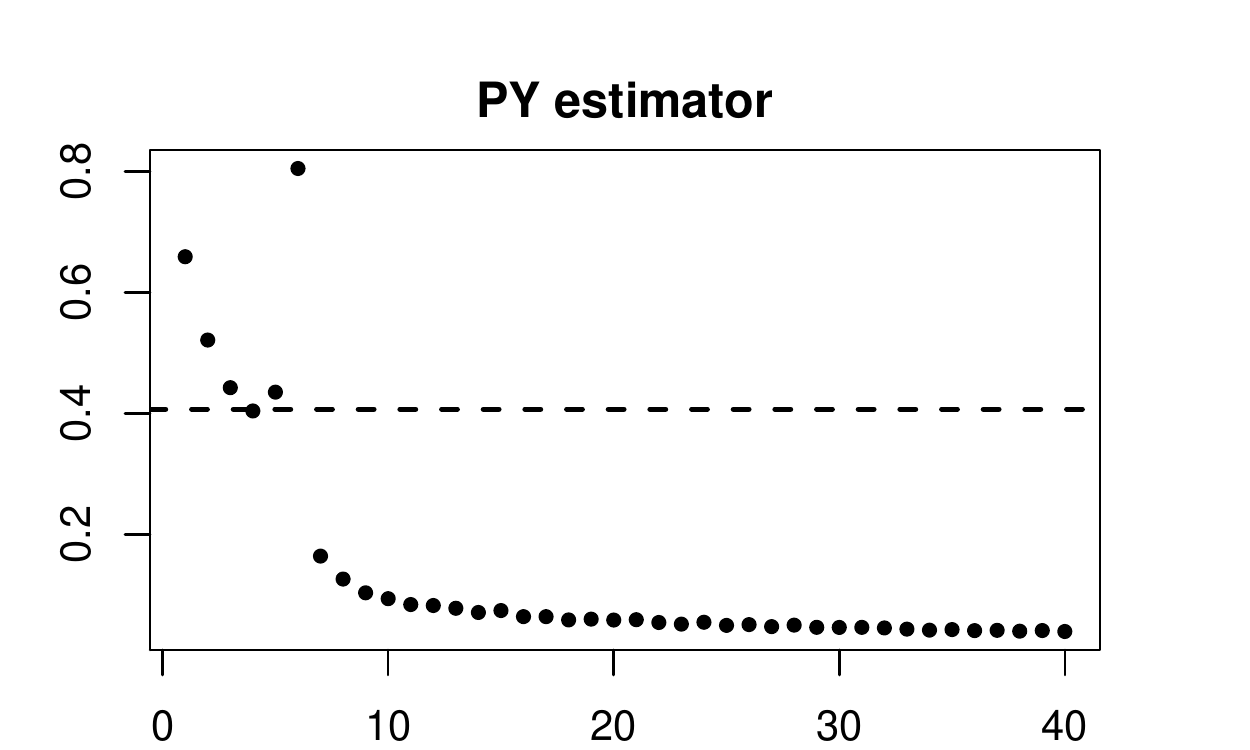}
\includegraphics[scale=0.5]{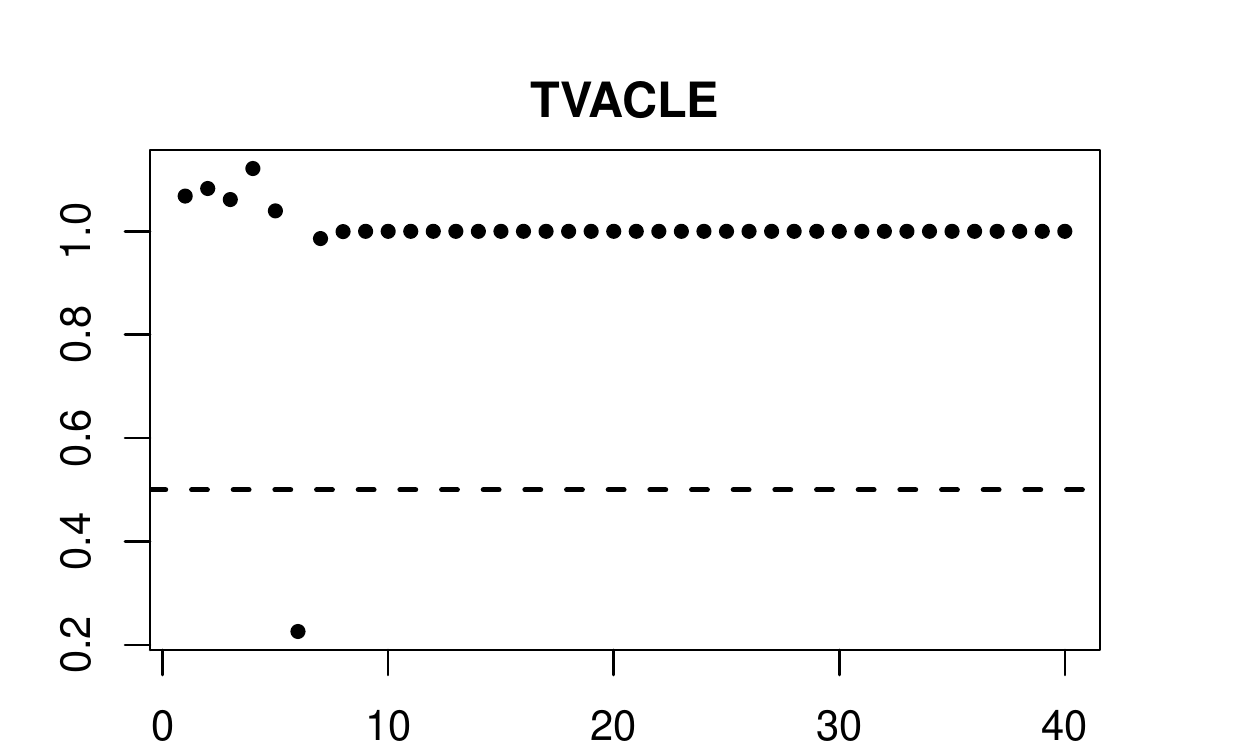}
\label{figure4}
\end{figure}

\begin{table}[H]\footnotesize
\caption{\footnotesize Mean, mean squared error and empirical distribution of $\hat{q}_n^{PY}$ and $\hat{q}_n^{TVACLE}$ over 500 independent replications for Model~\ref{model4} $(q=6)$, with known $\sigma^2=1$.}
\centering
{\small\scriptsize\hspace{12.5cm}
\renewcommand{\arraystretch}{1}\tabcolsep 0.1cm
\renewcommand\arraystretch{0.8}
\begin{tabular}{c|c|c|c|cccccccc}
\hline
&$(p,n$)&Mean&MSE&$\hat q=0$ &$\hat q=1$ &$\hat q=2$ &$\hat q=3$ &$\hat q=4$ &$\hat q=5$&$\hat q=6$&$\hat q\ge 7$\\
\hline
\multirow{6}{*}{$\hat{q}_n^{PY}$}
&$(50,200)$  &5.358&2.874&0.018&0.04&0.042&0.06&0&0&\textbf{0.826}&0.014\\
&$(200,800)$  &5.816&0.868&0.002&0.014&0.02&0.012&0&0&\textbf{0.94}&0.012\\
%\hline
%$(50,100)$  &5.208  &3.032 &0.020 &0.034 &0.050 &0.056 &0.004 &0.134 &\textbf{0.694} &0.008\\
%$(200,400)$  &5.788  &1.24   &0.016 &0.026  &0.0 &0.0 &0.0 &0.0 &\textbf{0.944} & 0.014\\
\cline{2-12}
&$(100,100)$  &4.436&6.904&0.06&0.072&0.118&0.106&0.01&0.048&\textbf{0.572}&0.014\\
&$(200,200)$  &4.964&4.772&0.042&0.052&0.082&0.07&0&0&\textbf{0.742}&0.012\\
\cline{2-12}
&$(400,200)$  &3.858&9.794&0.078&0.138&0.164&0.094&0.008&0.032&\textbf{0.484}&0.002\\
&$(800,400)$  &4.406&7.558&0.068&0.11&0.098&0.086&0&0&\textbf{0.626}&0.012\\
\hline
\multirow{6}{*}{$\hat{q}_n^{TVACLE}$}
&$(50,200)$  &6.006&0.006&0&0&0&0&0&0&\textbf{0.994}&0.006\\
&$(200,800)$  &6.024&0&0&0&0&0&0&0&\textbf{0.976}&0.024\\
\cline{2-12}
%$(50,100)$  &5.878 &0.122 &0.0 &0.0 &0.0 &0.0 &0.002 &0.122 &\textbf{0.878} &0.0\\
%$(200,400)$  &6.034 &0.034 &0.0 &0.0 &0.0 &0.0 &0.0 &0.0 &\textbf{0.966} &0.034\\
%\hline
&$(100,100)$  &5.886&0.122&0&0&0&0&0.004&0.106&\textbf{0.89}&0.11\\
&$(200,200)$  &6&0&0&0&0&0&0&0&\textbf{1}&0\\
\cline{2-12}
&$(400,200)$  &5.952&0.06&0&0&0&0&0.004&0.042&\textbf{0.952}&0.002\\
&$(800,400)$  &6.002&0.002&0&0&0&0&0&0&\textbf{0.998}&0.002\\
\hline
\end{tabular}
}
\label{table2}
\end{table}
%\begin{center}
%Table~\ref{table2} %and Figures~\ref{figure4}-\ref{figure7}
% about here
%\end{center}
To further confirm this phenomenon, we report the results for Model~\ref{model4} with more equal spikes. The results in Table~\ref{table2} suggest that  $\hat{q}_n^{TVACLE}$ overall performs better than $\hat{q}_n^{PY}$ in terms of estimation accuracy and MSE, It has underestimation problem as  its searching procedure stops earlier once the difference between consecutive eigenvalues corresponding to equal spikes are below the threshold $d_n$.
This conclusion can be made after observing its empirical distributions in Table~\ref{table2}. In contrast, $\hat{q}_n^{TVACLE}$ largely avoids this problem. To better illustrate this fact, we plot in Figure~\ref{figure4} the  first 40 differences $\hat\delta_i$ for  $\hat q_n^{PY}$
and the first 40 ratios of  $\hat r_i^{TR}$ for $\hat q_n^{TVACLE}$.
%This figure is based on simulations with 500 independent replications for Model~\ref{model4}, and $(p,n)=(400,200)$.
%It can be seen from 
The left subfigure shows that there are three $\hat \delta_i$, $i=3,4,5$, are very close to the threshold line $y=d_n$, which causes the underestimation problem shown in Table~\ref{table2}. In contrast,  the right subfigure shows that the ``valley'' $\hat r_q^{TR}$ and the ``cliff'' $\hat r_{q+1}^{TR}$ are well separated by the threshold line $\tau=0.5$.

\begin{figure}[H]
\centering
\caption{\footnotesize Boxplots of the first 7 ratios: the left is for ratios without ridge, $\hat r_{i}$; the middle is for ratios with ridge $\hat r_{i}^R$; the right is for transformed ratios with ridge $\hat r_{i}^{TR}$.}
\includegraphics[scale=0.5]{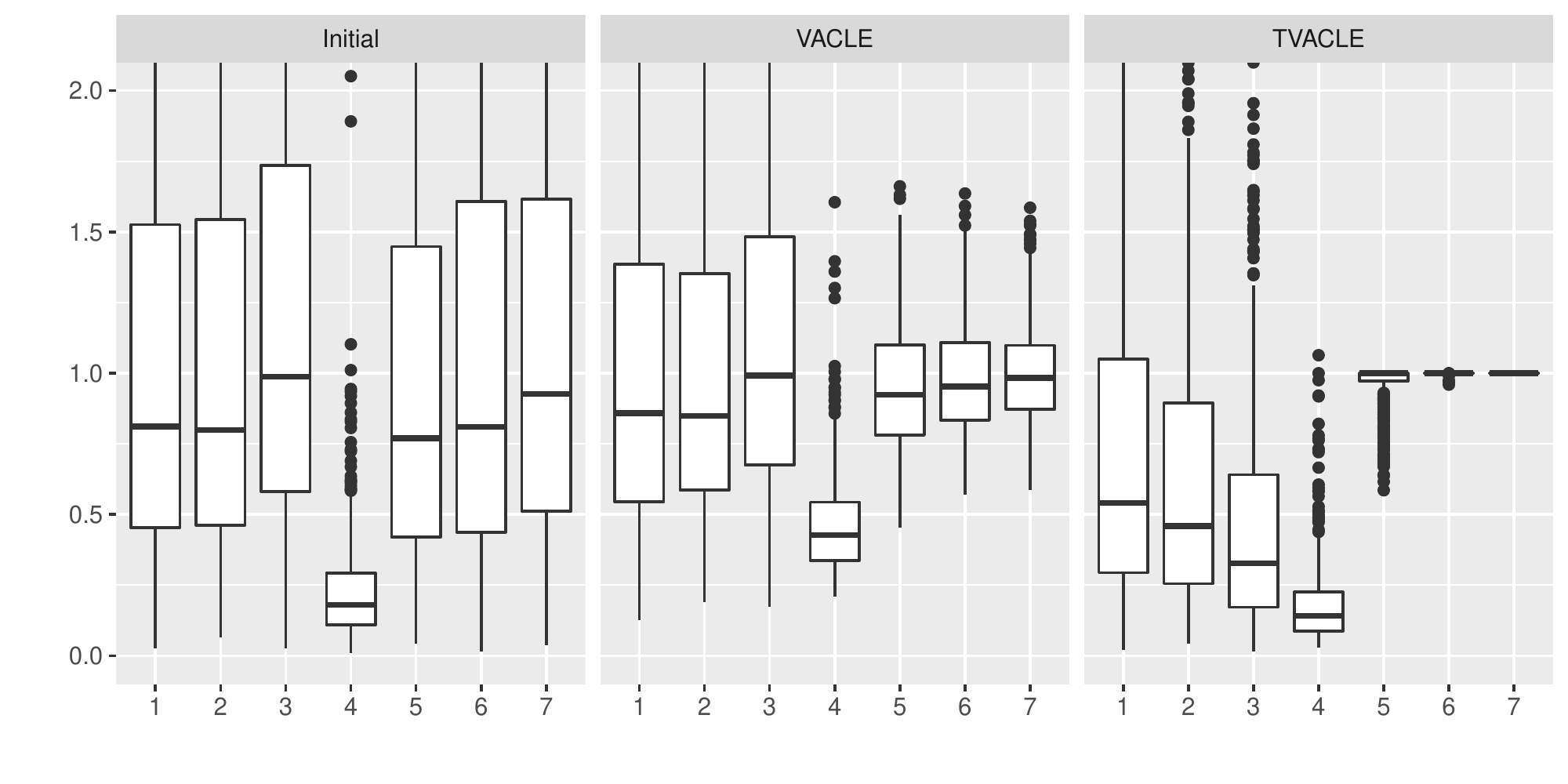}
\label{figure1}
\end{figure}

As we claimed in Sections~\ref{sec2} and \ref{sec3}, the VACLE could be somehow sensitive to the ridge selection. The results reported in Table~\ref{table1} confirm this claim. %The TVACLE shows better performance than the VACLE in this simulation. 
To explore the manner how the ridge $c_n$ affects both the VACLE and the TVACLE, Figures~\ref{figure1} presents, for Model~\ref{model2} with $(p,n)=(400,200)$, the boxplots of the first 7 ratios without ridge $\hat r_{i}$; the first 7 ridge ratios $\hat r_{i}^R$; and the first 7 transformed ridge ratios  $\hat r_{i}^{TR}$.
 From the left to right  subfigure of Figure~\ref{figure1}, we can see that   for $i>q=4$, $\hat r_{i}$ fluctuates much more than $\hat r_{i}^R$ and $\hat r_{4}^{TR}$ and $\hat r_{i}^{TR}$, $i>4$ are separated more significantly. This confirms the necessity of using a ridge with a stabilised ratio $\hat r_{i}^R$ and transformation can enhance the estimation accuracy. 

%\vspace{-0.8em}
%\begin{figure}
%\centering
%\includegraphics[scale=0.45]{4.png}
%\caption{The frequency of correct estimation for $\hat q_n^{PY}$ and $\hat q_n^{TVACLE}$, in Model 4 and $c=0.25$.}
%\label{figure4}
%\end{figure}

%\begin{figure}
%\centering
%\includegraphics[scale=0.45]{5.png}
%\caption{The frequency of correct estimation for $\hat q_n^{PY}$ and $\hat q_n^{TVACLE}$, in Model 4 and $c=0.5$.}
%\label{figure5}
%\end{figure}
%
%\begin{figure}
%\centering
%\includegraphics[scale=0.45]{6.png}
%\caption{The frequency of correct estimation for $\hat q_n^{PY}$ and $\hat q_n^{TVACLE}$, in Model 4 and $c=1$.}
%\label{figure6}
%\end{figure}
%
%\begin{figure}
%\centering
%\includegraphics[scale=0.45]{7.png}
%\caption{The frequency of correct estimation for $\hat q_n^{PY}$ and $\hat q_n^{TVACLE}$, in Model 4 and $c=2$.}
%\label{figure7}
%\end{figure}

\subsubsection*{The unknown $\sigma^2$ Case.}
Use Models~\ref{model2} and \ref{model4} and  regard $\sigma^2$ as an unknown value. These two models represent the cases with and without equal spikes. Furthermore, because the conclusions are very similar to those with known $\sigma^2$, we then report only the results for $\hat{q}_n^{VACLE}$ and $\hat{q}_n^{TVACLE}$ to further confirm the advantages of $\hat{q}_n^{TVACLE}$. The numerical results are shown in Table~\ref{table3}. The results in the last two columns show  that the one-step estimation $\hat\sigma^2$ has good performance in terms of accuracy and robustness.
 %In addition, we also report the mean and mean square error of $\hat{\sigma}^2$ to show the performance of our estimation procedure.
%In addition, the $\hat\sigma^2$ works  well in terms of consistency and robustness. This should be the reason that there does not exist significant difference between the known and unknown $\sigma^2$ cases.

\begin{table}[h]\footnotesize
\caption{\footnotesize Mean and mean square error of $\hat{q}_n^{VACLE}$, $\hat{q}_n^{TVACLE}$ and $\hat\sigma^2$, and the misestimation rates of  $\hat{q}_n^{VACLE}$ and $\hat{q}_n^{TVACLE}$ over 500 independent replications for Model \ref{model2} and \ref{model4}, with unknown $\sigma^2$ whose true value is 1.}
\centering
{\small\scriptsize\hspace{12.5cm}
\renewcommand{\arraystretch}{1}\tabcolsep 0.1cm
\renewcommand\arraystretch{0.8}
\begin{tabular}{c|c|ccc|ccc|cc}
\hline
& & \multicolumn{3}{c}{$\hat{q}_n^{VACLE}$} &\multicolumn{3}{c}{$\hat{q}_n^{TVACLE}$}&\multicolumn{2}{c}{$\hat\sigma^2$} \\
&$(p,n)$& \multicolumn{1}{c}{Mean}&\multicolumn{1}{c}{MSE}&\multicolumn{1}{c}{$\hat{q}_n^{VACLE}\neq q$}&\multicolumn{1}{c}{Mean}&\multicolumn{1}{c}{MSE} &\multicolumn{1}{c}{$\hat{q}_n^{TVACLE}\neq q$} &\multicolumn{1}{c}{Mean}&\multicolumn{1}{c}{MSE}\\
\hline
\multirow{6}{*}{Model \ref{model2}}
&$(50,200)$     &4.002 &0.002 &0.002 &4.012 &0.012 &0.012 &1.0513 &0.0033 \\
%$(100,400)$     &4  &0   &0  &4.028   &0.028 &0.028 &1.0237 &0.0007\\
&$(200,800)$    &4.002 &0.002 &0.002 &4.014 &0.014 &0.014 &1.0119 &0.0002 \\
%$(400,1600)$     &4  &0   &0  &4.002   &0.002 &0.002 &1.0056 &$4\times 10^{-5}$\\
\cline{2-10}
%$(50,100)$   &3.972 &0.284 &0.062 &4.01 &0.01 &0.01 &1.0627 &0.0053\\
%$(100,200)$  &3.994 &0.038 &0.008 &4.046 &0.05 &0.044 &1.0282 &0.0012\\
%$(200,400)$   &4 &0 &0 &4.044 &0.044 &0.044 &1.0150 &0.0003\\
%$(400,800)$  &4 &0 &0 &4.004 &0.004 &0.004 &1.0074 &$8\times 10^{-5}$\\
%\hline
%$(50,50)$   &3.268 &3.312 &0.502 &3.804 &0.212 &0.212 &1.0741 &0.0094\\
&$(100,100)$  &3.326 &2.346 &0.258 &3.966 &0.038 &0.038 &1.0326 &0.0022 \\
&$(200,200)$ &3.96 &0.176 &0.04 &4.006 &0.006 &0.006 &1.0169 &0.0006 \\
%$(400,400)$  &4.002 &0.002 &0.002 &4.058 &0.066 &0.054 &1.0076 &0.0001\\
\cline{2-10}
%$(50,25)$  &3.01 &6.022 &0.802 &2.89 &1.562 &0.886 &1.0891 &0.0129\\
%&$(100,50)$  &3.076 &5.424 &0.744 &3.452 &0.592 &0.532 &1.0395 &0.0029\\
&$(200,100)$  &2.334 &5.726 &0.616 &3.71 &0.306 &0.282 &1.0205 &0.0008 \\
&$(400,200)$ &3.266 &2.454 &0.292 &3.962 &0.038 &0.038 &1.0094 &0.0002 \\
\hline
\multirow{6}{*}{Model \ref{model4}}
&$(50,200)$  &6.01 &0.01 &0.01 &6.01 &0.01 &0.01 &1.0788 &0.0069 \\
%$(100,400)$   &6.002 &0.002 &0.002 &6.032 &0.032 &0.032 &1.0359 &0.0015\\
&$(200,800)$   &6.002 &0.002 &0.002 &6.022 &0.022 &0.022 &1.0181 &0.0004 \\
%$(400,1600)$   &6 &0 &0 &6 &0 &0 &1.0087 &$9\times 10^{-5}$\\
\cline{2-10}
%$(50,100)$  &4.492 &7.464 &0.462 &5.856 &0.156 &0.15 &1.0958 &0.0107\\
%$(100,200)$  &5.282 &3.706 &0.186 &6.012 &0.036 &0.036 &1.0443 &0.0023\\
%$(200,400)$  &5.94 &0.264 &0.024 &6.02 &0.02 &0.02 &1.0204 &0.0005\\
%$(400,800)$  &6 &0 &0 &6.01 &0.014 &0.008 &1.0108 &0.0001\\
%\hline
&$(100,100)$  &4.082 &10.362 &0.388 &5.878 &0.142 &0.112 &1.0555 &0.0042 \\
&$(200,200)$ &5.846 &0.938 &0.034 &6 &0 &0 &1.0256 &0.0009 \\
%&$(400,400)$  & & & & & & & & \\
%$(800,800)$  &5.684 &1.464 &0.118 &6.01 &0.018 &0.018 &1.0065 &$6\times 10^{-5}$\\
\cline{2-10}
%$(200,100)$  &2.786 &16.23 &0.912 &4.786 &1.85 &0.922 &1.0319 &0.0014\\
&$(400,200)$  &4.524 &8.064 &0.306 &5.958 &0.042 &0.042 &1.0165 &0.0004 \\
&$(800,400)$  &5.822 &0.966 &0.034 &6.01 &0.01 &0.01 &1.0079 &$9\times 10^{-5}$ \\
%$(1600,800)$ &3.812 &9.988 &0.822 &5.548 &0.476 &0.47 &1.0037 &$2\times 10^{-5}$\\
\hline
\end{tabular}
}
\label{table3}
\end{table}

\subsection{ Numerical studies on large dimensional auto-covariance matrix}
To estimate the number of factors in Model (\ref{auto-model}), \cite{li2017identifying} introduced the following ratio-based estimator,
\begin{equation}\label{LWY estimator}
\hat q_T^{LWY}=\min\{i\ge 1:\ {\hat\lambda_{i+1}}/{\hat\lambda_{i}}>1-d_T\ \text{and}\ {\hat\lambda_{i+2}}/{\hat\lambda_{i+1}}>1-d_T\}-1,
\end{equation}
where $\hat\lambda_{i},1\le i\le p$, are in descending order and $d_T$ is a tuning parameter selected as that in  Section 3.1 of  \cite{li2017identifying}. %In addition, their simulations showed that this method has overall better performance than the estimator proposed by \cite{lam2012factor}. Thus, 
To examine the performance of the $\hat{q}_n^{TVACLE}$, we use $\hat q_T^{LWY}$ as the competitor. For the ratio
%between the dimension and sample size
$p/T=y$, we consider two values $y=0.5$ and $y=2$. The dimension $p=100,200,300,400\ \text{and}\ 500$. In each case, we repeat the experiment 500 times. To be fair and concise, we conduct the simulation with two models as follows. The model structure is the same as in \cite{lam2012factor} and \cite{li2017identifying}: for $\ 1\le t\le T$
\begin{equation}
y_t=\mathbf{A}x_t+\varepsilon_t,\ \varepsilon_t\sim \mathbb{N}_p(\textbf{0},\textbf{I}_p),\ x_t=\Theta x_{t-1}+e_t,\ e_t\sim \mathbb{N}_k(\textbf{0},\textbf{$\Gamma$}),
\end{equation}
where $\mathbf{A}\in \mathbb{R}^{p\times q}$ is the factor loading matrix and $\{\varepsilon_t\}$ is a white noise sequence with unit variance $\sigma^2=1$. As in \cite{li2017identifying}, $\mathbf{A}$ and $\Gamma$ take the forms as $\mathbf{A}=(\mathbf{I}_q, \mathbf{O}_{(p-q)\times q})^{\top}$, $\Gamma=diag(2,2,\cdots,2)$.
%\begin{eqnarray}
%\mathbf{A}=(\mathbf{I}_q, \mathbf{O}_{(p-q)\times q})',\
%\Gamma=diag(2,2,\cdots,2).
%\end{eqnarray}
We manipulate the strength of factors by adjusting the matrix $\Theta$ in different models as follows:
\begin{model}\label{model5}
This model is the same as \emph{Scenario III} in \cite{li2017identifying}. There are $q=3$ factors whose theoretical limits
%$(\lambda_1,\lambda_2,\lambda_3)$
equal $(7.726,5.496,3.613)$ in the case of $y=0.5$ and $(23.744,20.464,17.970)$ in the case of $y=2$. The upper edge $b_1$ of the supports in these two cases are respectively 2.773 and 17.637.  $q=3$  factors are identifiable, and $\Theta=diag(0.6,-0.5,0.3)$.
%\begin{eqnarray}
%\Theta=diag(0.6,-0.5,0.3).
%\end{eqnarray}
\end{model}
\begin{model}\label{model6}
This model has more factors. There are $q=6$ factors with identical strength, and their theoretical limits are 5.496 in the case of $y=0.5$ and $20.464$ in the case of $y=2$. Because these limits exceed their corresponding upper edge $b_1$, all $q=6$ factors are identifiable in theory with $\Theta=diag(0.5,0.5,0.5,0.5,0.5,0.5)$.
%\begin{eqnarray}
%\Theta=diag(0.5,0.5,0.5,0.5,0.5,0.5).
%\end{eqnarray}
\end{model}
All parameters in the simulations
%for auto-covariance matrix
share the same settings of parameters  in Section~\ref{sec5.2} where we conduct numerical studies for spiked population models. These parameters in TVACLE estimator are shown in Table~\ref{auto parameter}.
\begin{table}[h]\footnotesize
%\caption{Parameters for TVACLE.}
\centering
\caption{\footnotesize Parameters in the TVACLE estimator.}
\renewcommand\arraystretch{0.8}
\begin{tabular}{cccccc}
\toprule
 $\tau$& $c_T$ & $\kappa_T$ &$k_1$ &$k_2$ &$L$\\
\midrule
0.5 &$\sqrt{\log\log T}\cdot[\text{q}_{p,T}(0.95)-\text{q}_{p,T}(0.05)]-\text{m}_{p,T}$&$\log\log p \cdot p^{-2/3}$&5&5&20\\
\bottomrule
\end{tabular}\label{auto parameter}
%\label{table}
\end{table}
%Recalling the aforementioned trade-off between underestimation and overestimation for ridge selection, $c_n$ should be larger than those $\hat\delta_{T,i}$ for $i>q$, but not too much. Hence, we consider one kind of ridge based on the sample median of $\hat\delta_{T,q+1}$. For given $(p,T)$, we conduct 500 times independent pure-noise simulations and then get the sample median of the difference sequence $\{\hat l_1-\hat l_2\}$, say $m_{p,T}$, where $\hat l_1$ and $\hat l_2$ are the two largest eigenvalue of the noise matrix $\hat M_{\varepsilon}$. Using $m_{p,T}$ to approximate the sample median of $\hat\delta_{T,q+1}=\hat \lambda_{T,q+1}-\hat \lambda_{T,q+2}$ and multiplying it by $\log\log p$, we get the ridge $c_n$ above. Note that $m_{p,T}$ converges to zero at the same rate as $\hat \lambda_{T,q+1}$ that has a slightly faster rate to zero than $c_T$. The results are reported in Tables~\ref{table6} and \ref{table7}.

\begin{table}[H]\footnotesize
\caption{\footnotesize Mean, mean squared error and empirical distribution of $\hat{q}_T^{LWY}$ and $\hat{q}_T^{TVACLE}$ over 500 independent replications for Model \ref{model5}.}
\centering
{\small\scriptsize\hspace{12.5cm}
\renewcommand\arraystretch{0.8}
\begin{tabular}{c|cccccc|cccccc}
\hline
&$p$ &100 &200 &300 &400 &500 &$p$ &100 &200 &300 &400 &500 \\
&$T=2p$ &200 &400 &600 &800 &1000 &$T=0.5p$ &50 &100 &150 &200 &250\\
\hline
\multirow{7}{*}{$\hat{q}_T^{LWY}$}
&$\hat q= 0$  &0.024&0.002&0&0&0&$\hat q= 0$  &0.53&0.238&0.234&0.138&0.054\\
&$\hat q= 1$  &0.028&0&0&0&0&$\hat q=1$  &0.326&0.412&0.38&0.36&0.282\\
&$\hat q = 2$  &0.384&0.138&0.05&0.014&0.008&$\hat q= 2$  &\textbf{0.136} &\textbf{0.32} &\textbf{0.356} &\textbf{0.464} &\textbf{0.572}\\
&$\hat q = 3$  &\textbf{0.544} &\textbf{0.85} &\textbf{0.948} &\textbf{0.976} &\textbf{0.986} &$\hat q= 3$  &\textbf{0.008} &\textbf{0.03} &\textbf{0.03} &\textbf{0.036} &\textbf{0.092}\\
&$\hat q \ge 4$ &0.02&0.01&0.002&0.01&0.006&$\hat q\ge 4$  &0&0&0&0.002&0\\
&Mean  &2.508&2.866&2.952&2.996&2.998& Mean  &0.622&1.142&1.182&1.404&1.702\\
&MSE  &0.732&0.166&0.052&0.024&0.014&MSE  &6.21&4.11&3.982&3.148&2.186\\
\hline
\multirow{7}{*}{$\hat{q}_T^{TVACLE}$}
&$\hat q=0$  &0&0&0&0&0&$\hat q= 0$  &0.02&0.002&0&0.002&0\\
&$\hat q= 1$  &0&0&0&0&0&$\hat q= 1$  &0.332&0.182&0.116&0.104&0.054\\
&$\hat q = 2$  &0.196&0.02&0.014&0.008&0.002&$\hat q= 2$  &\textbf{0.584} &\textbf{0.698} &\textbf{0.688} &\textbf{0.73} &\textbf{0.676}\\
&$\hat q = 3$  &\textbf{0.782} &\textbf{0.948} &\textbf{0.974} &\textbf{0.964} &\textbf{0.974} &$\hat q= 3$  &\textbf{0.062} &\textbf{0.116} &\textbf{0.196} &\textbf{0.16} &\textbf{0.268}\\
&$\hat q \ge 4$  &0.022&0.032&0.012&0.028&0.024 &$\hat q\ge 4$  &0.002&0.002&0&0.004&0.002\\
&Mean  &2.826&3.012&2.998&3.02&3.022 &Mean  &1.694&1.934&2.08&2.06&2.218\\
&MSE  &0.218&0.052&0.026&0.036&0.026 &MSE  &2.094&1.446&1.152&1.168&0.894\\
\hline
\end{tabular}
}
\label{table6}
\end{table}

\begin{table}[H]\footnotesize
\caption{\footnotesize Mean, mean squared error and empirical distribution of $\hat{q}_T^{LWY}$ and $\hat{q}_T^{TVACLE}$ over 500 independent replications for Model \ref{model6}.}
\centering
{\small\scriptsize\hspace{12.5cm}
\renewcommand\arraystretch{0.8}
\begin{tabular}{c|cccccc|cccccc}
\hline
&$p$ &100 &200 &300 &400 &500 &$p$ &100 &200 &300 &400 &500 \\
&$T=2p$ &200 &400 &600 &800 &1000 &$T=0.5p$ &50 &100 &150 &200 &250\\
\hline
\multirow{10}{*}{$\hat{q}_T^{LWY}$}
&$\hat q = 0$  &0.156&0.104&0.072&0.098&0.054 &$\hat q= 0$  &0.226&0.202&0.26&0.24&0.134\\
&$\hat q = 1$  &0.178&0.154&0.124&0.146&0.076 &$\hat q= 1$  &0.418&0.35&0.326&0.304&0.28\\
&$\hat q = 2$  &0.19&0.154&0.114&0.104&0.062 &$\hat q= 2$  &0.296&0.314&0.262&0.236&0.312\\
&$\hat q = 3$  &0.162&0.112&0.068&0.106&0.044 &$\hat q= 3$  &0.06&0.122&0.134&0.17&0.188\\
&$\hat q = 4$  &0.13&0.006&0&0&0 &$\hat q= 4$  &0&0.012&0.016&0.05&0.07\\
&$\hat q = 5$  &0.112&0.072&0&0&0 &$\hat q= 5$  &0&0&0.002&0&0.014\\
&$\hat q = 6$  &\textbf{0.072} &\textbf{0.394} &\textbf{0.62} &\textbf{0.542} &\textbf{0.754} &$\hat q= 6$  &\textbf{0} &\textbf{0} &\textbf{0} &\textbf{0} &\textbf{0.002}\\
&$\hat q \ge 7 $  &0&0.004&0.002&0.004&0.01 &$\hat q\ge 7$  &0&0&0&0&0\\
&Mean  &2.556&3.574&4.29&3.954&4.926 &Mean  &1.19&1.392&1.326&1.486&1.83\\
&MSE  &15.196&11.166&8.13&9.806&5.242 &MSE  &23.862&22.192&22.974&21.746&18.802\\
\hline
\multirow{10}{*}{$\hat{q}_T^{TVACLE}$}
&$\hat q = 0$  &0 &0 &0 &0 &0 &$\hat q= 0$  &0 &0 &0 &0 &0\\
&$\hat q = 1$  &0 &0 &0 &0 &0 &$\hat q= 1$  &0.01 &0 &0 &0 &0\\
&$\hat q = 2$  & 0&0 &0 &0 &0 &$\hat q= 2$  &0.206&0.07&0.03&0.008&0.008\\
&$\hat q = 3$  & 0.004&0 &0 &0 &0 &$\hat q= 3$  &0.586&0.496&0.33&0.224&0.13\\
&$\hat q = 4$  &0.066&0.002&0&0&0 &$\hat q= 4$  &0.19&0.414&0.546&0.574&0.554\\
&$\hat q = 5$  &0.418&0.038&0&0&0 &$\hat q= 5$  &0.008&0.02&0.094&0.188&0.294\\
&$\hat q = 6$  & \textbf{0.51}&\textbf{0.946} &\textbf{0.99} &\textbf{0.984} &\textbf{0.97} &$\hat q= 6$  &\textbf{0} &\textbf{0} &\textbf{0} &\textbf{0.006} &\textbf{0.014}\\
&$\hat q \ge 7 $  &0.002&0.014&0.01&0.016&0.03 &$\hat q\ge 7$  &0 &0 &0 &0 &0\\
&Mean  &5.44&5.972&6.01&6.016&6.03 &Mean  &2.98&3.384&3.704&3.96&4.176\\
&MSE  &0.72&0.06&0.01&0.016&0.03 &MSE  &9.588&7.26&5.728&4.628&3.808\\
\hline
\end{tabular}
}
\label{table7}
\end{table}
From Table~\ref{table6}, we can see that when $T=2p$, $\hat q_T^{LWY}$ works well. That is, when $T$ is large, $\hat q_T^{LWY}$ shows good performance, whilst when $T$ is not large, it tends to underestimate the true number $q$. Our method clearly outperforms $\hat q_T^{LWY}$. Although when $T$ is small, the true value is somewhat underestimated, but still, with high proportion, to be two or greater. Table~\ref{table7} shows that for Model~\ref{model6} with equal spikes, when $T=2p$, the performance of $\hat q_T^{LWY}$ is not encouraging, and when $T=0.5p$, the underestimation problem becomes very serious, with a very high proportion having $\hat q_T^{LWY} \le 2$. In contrast, our method performs well when $T=2p$ and when $T=0.5p$; underestimation still occurs, but it is much less serious than  $\hat q_T^{LWY}$  in the sense that $\hat q> 2$ with high proportion. Overall, our estimator $\hat{q}_T^{TVACLE}$ is  superior to $\hat q_T^{LWY}$ in these limited simulations.

\subsection{Numerical studies on large dimensional spiked Fisher matrix}
Because the TVACLE has been demonstrated to outperform the VACLE overall, we only consider the comparison between $\hat q_n^{TVACLE}$ and the estimator $\hat q_n^{WY}$ introduced by \cite{wang2017extreme}. Sharing notations in Section~\ref{sec4}, the
%scree plot based
estimator $\hat q_n^{WY}$ can be written as
\begin{equation}
\hat q_n^{WY}=\max\{i:\hat\lambda_{i}\ge b_2+d_n\},
\end{equation}
where $d_n$ was recommended to be $(\log\log p)p^{-2/3}$ in their paper.

As a Fisher matrix $\textbf{F}_n=\textbf{S}_1\textbf{S}_2^{-1}$ involves two random matrices $\textbf{S}_1$ and $\textbf{S}_2$,  its eigenvalues are more dispersed, with wider range of the support, % , and these extreme eigenvalues have much more fluctuation
than the spiked sample covariance matrices and auto-covariance matrices. %This fact can be found by examining the range of its support. 
The aforementioned automatic procedure for ridge selection  would then generate a larger $c_n$, and this in turn increases the value at the ``valley''. Hence, %to search for the value at the ``valley'' right before the ``cliff'', 
we use a larger threshold $\tau=0.8$  to avoid underestimation. Further, in the following Model~\ref{model7}, we set the ridge $c_n^{(3)}=\sqrt{\log\log p}\cdot[\text{q}_{p,n}(0.95)-\text{q}_{p,n}(0.05)]-\text{m}_{p,n}$, whilst for  Model~\ref{model8} with dramatically-fluctuated extreme eigenvalues, we need to set $c_n^{(3)}=\sqrt{\log\log p}\cdot[\text{q}_{p,n}(0.8)-\text{q}_{p,n}(0.05)]-\text{m}_{p,n}$ to avoid too large ridge. The parameters in $\hat q_n^{TVACLE}$ are shown in Table~\ref{Fisher parameters}.

\begin{table}[h]\footnotesize
\caption{\footnotesize Parameters in the TVACLE estimator.}
\centering
\renewcommand\arraystretch{0.8}
\begin{tabular}{cccccc}
\toprule
 $\tau$& $c_n$ & $\kappa_n$ &$k_1$ &$k_2$ &$L$\\
\midrule
   0.8 &$c_n^{(3)}$&$\log\log p\cdot p^{-2/3}$&5&5&20\\
\bottomrule
\end{tabular}
\label{Fisher parameters}
\end{table}

Again, for a fair comparison, we design two models, one that was used by \cite{wang2017extreme} and the other with weaker spikes. For $y=p/T$ and $c=p/n$,
%the two limits $y=p/T$ and $c=p/n$,
we set $(0.5,0.2)$ and $(0.2,0.5)$ for the respective models. The dimension $p$ takes values of $50, 100, 150, 200$ and $250$. For each combination $(p,T,n)$, the experiment is repeated $500$ times. Consider the number of spikes to be $q=3$ and $\textbf{A}$ to be a $p\times 3$ matrix as:
\begin{eqnarray}
\left(
  \begin{array}{cccccc}
    \sqrt {\alpha_1} & 0 & 0 & 0 & \cdots & 0 \\
    0 & \sqrt{\frac{\alpha_2}{2}} & \sqrt{\frac{\alpha_2}{2}} & 0 & \cdots & 0 \\
    0 & \sqrt{\frac{\alpha_3}{2}} & -\sqrt{\frac{\alpha_3}{2}} & 0 & \cdots & 0 \\
  \end{array}
\right)^T_{3\times p},
\end{eqnarray}
where $\alpha=(\alpha_1,\alpha_2,\alpha_3)$ assumes different values in two models. Assume the covariance matrix $Cov(u_i)=\textbf{I}_3$ and $\Sigma_2=diag(1,\cdots,1,2,\cdots,2)$, where ``1'' and ``2'' both have multiplicity $p/2$. The two models are:
\begin{model}\label{model7}
Let $\alpha=(10,5,5)$, $(y,c)=(0.5,0.2)$, which is Model 1 in \cite{wang2017extreme}. The matrix $\Sigma_1\Sigma_2^{-1}$ has three spikes $\lambda_1=11$ and $\lambda_2=\lambda_2=6$ that are all significantly larger than the upper bound $b_2=\frac{1+\sqrt{c+y-cy}}{1-y}\approx~3.55$ of  support of the distribution. %Because all three eigenvalues are significantly larger than the upper bound, there are three strong detectable signals.
\end{model}
\begin{model}\label{model8}
Let $\alpha=(10,2,2)$, $(y,c)=(0.2,0.5)$. The matrix $\Sigma_1\Sigma_2^{-1}$ then also has three spikes $\lambda_1=11$ and $\lambda_2=\lambda_2=3$ larger than the upper bound $b_2=\frac{1+\sqrt{c+y-cy}}{1-y}\approx~2.22$ of the support of the distribution. Then $\lambda_2=\lambda_2=3$ are relatively more difficult to detect.
\end{model}

\begin{table}[H]\footnotesize
\caption{\footnotesize Mean, mean squared error and empirical distribution of $\hat{q}_n^{WY}$ and $\hat{q}_n^{TVACLE}$ % over 500 independent replications 
for Model \ref{model7}.}
\centering
{\small\scriptsize\hspace{12.5cm}
\renewcommand\arraystretch{0.8}
\begin{tabular}{c|c|c|c|ccccc}
\hline
&$(p,T,n)$&Mean&MSE&$\hat q=0$ &$\hat q=1$ &$\hat q=2$ &$\hat q=3$ &$\hat q=4$\\
\hline
&$(50,100,250)$  &2.344 &0.732 &0 &0.034 &0.592 &\textbf{0.37} &0.004\\
&$(100,200,500)$  &2.672 &0.352 &0 &0.004 &0.328 &\textbf{0.66} &0.008\\
$\hat{q}_n^{WY}$& $(150,300,750)$  &2.822 &0.194 &0 &0 &0.186 &\textbf{0.806} &0.008\\
&$(200,400,1000)$ &2.964 &0.092 &0 &0 &0.064 &\textbf{0.908} &0.028\\
&$(250,500,1250)$  &2.96 &0.068 &0 &0 &0.054 &\textbf{0.932} &0.014\\
\hline
&$(50,100,250)$  &2.364 &0.7 &0 &0.028 &0.584 &\textbf{0.384} &0.004\\
&$(100,200,500)$  &2.688 &0.336 &0 &0.004 &0.312 &\textbf{0.676} &0.008\\
$\hat{q}_n^{TVACLE}$ &$(150,300,750)$  &2.842 &0.182 &0 &0 &0.17 &\textbf{0.818} &0.012\\
&$(200,400,1000)$ &2.974 &0.082 &0 &0 &0.054 &\textbf{0.918} &0.028\\
&$(250,500,1250)$  &2.964 &0.064 &0 &0 &0.05 &\textbf{0.936} &0.014\\
\hline
\end{tabular}
}
\label{table4}
\end{table}

\begin{table}[H]\footnotesize
\caption{\footnotesize Mean, mean squared error and empirical distribution of $\hat{q}_n^{WY}$ and $\hat{q}_n^{TVACLE}$ % over 500 independent replications 
for Model \ref{model8}.}
\centering
{\small\scriptsize\hspace{12.5cm}
\renewcommand\arraystretch{0.8}
\begin{tabular}{c|c|c|c|ccccc}
\hline
&$(p,T,n)$&Mean&MSE&$\hat q=0$ &$\hat q=1$ &$\hat q=2$ &$\hat q=3$ &$\hat q=4$\\
\hline
&$(50,250,100)$  &2.114 &1.07 &0 &0.09 &0.708 &\textbf{0.2} &0.002\\
&$(100,500,200)$  &2.302 &0.79 &0 &0.046 &0.606 &\textbf{0.348} &0\\
$\hat{q}_n^{WY}$ &$(150,750,300)$  &2.498 &0.538 &0 &0.018 &0.466 &\textbf{0.516} &0\\
&$(200,1000,400)$ &2.622 &0.394 &0 &0.006 &0.368 &\textbf{0.624} &0.002\\
&$(250,1250,500)$  &2.692 &0.324 &0 &0.004 &0.304 &\textbf{0.688} &0.004\\
\hline
&$(50,250,100)$  &2.238 &0.898 &0 &0.064 &0.638 &\textbf{0.294} &0.004\\
&$(100,500,200)$  &2.462 &0.602 &0 &0.03 &0.48 &\textbf{0.488} &0.002\\
$\hat{q}_n^{TVACLE}$ &$(150,750,300)$&2.71 &0.314 &0 &0 &0.302 &\textbf{0.686} &0.012\\
&$(200,1000,400)$ &2.82 &0.232 &0 &0.002 &0.2 &\textbf{0.774} &0.024\\
&$(250,1250,500)$  &2.904 &0.164 &0 &0 &0.13 &\textbf{0.836} &0.034\\
\hline
\end{tabular}
}
\label{table5}
\end{table}

The results reported in Tables~\ref{table4} and \ref{table5} show that $\hat q_n^{TVACLE}$ shows overall better performance than $\hat q_n^{WY}$. % even for Model~7.
For Model~\ref{model8}, $\hat q_n^{TVACLE}$ is superior to $\hat q_n^{WY}$ when the signals are relatively weak. %Besides, consistency can be easily concluded for both $\hat q_n^{WY}$ and $\hat q_n^{TVACLE}$, which is also confirmed by Figure \ref{figure8} and \ref{figure9}.

\section{Real data example}\label{sec6}
Consider a data set of the daily prices of 100 stocks (see \cite{li2017identifying}).  This dataset includes the stock prices of the S\&P500 from 2005-01-03  to 2006-12-29. %to 2011-09-16.
Except for incomplete data, every stock has 502 observations of log returns. Thus, $T=502$, $p=100$, and then $c=p/T\approx0.2$.

\begin{figure}[h]
\centering
\caption{\footnotesize Eigenvalues of $\hat {\mathbf{M}}_y$ from $\hat\lambda_3$ to $\hat\lambda_{40}$}
\includegraphics[scale=0.5]{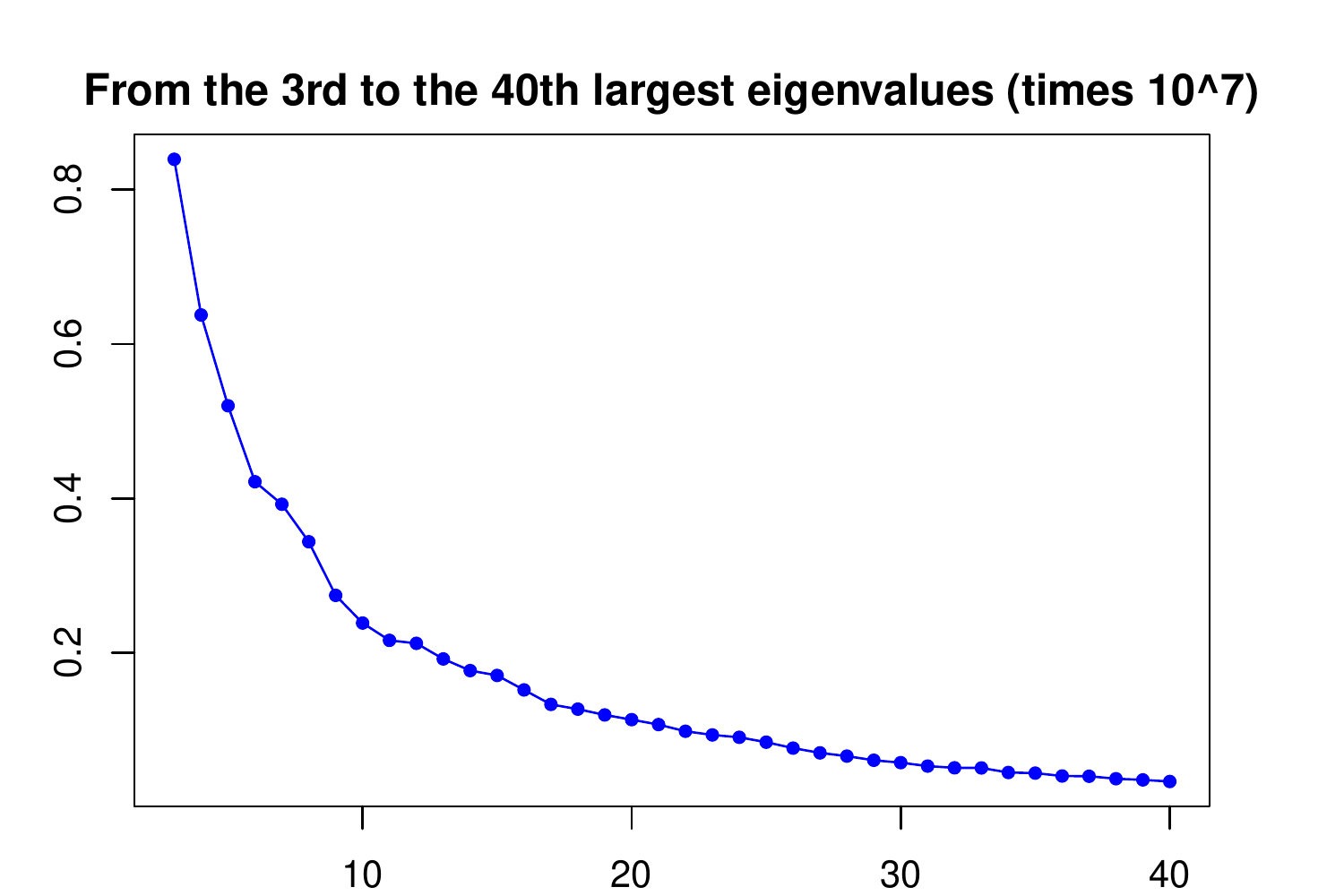}
\label{figure2}
\end{figure}

Denote $y_t\in \mathbb{R}^p$, $1\le t\le T$, as the $t$-th observation of the log return of these 100 stocks, and we then obtain its lag-1 sample auto-covariance matrix $\hat \Sigma_y$ and the matrix $\hat {\mathbf{M}}_y=\hat \Sigma_y\hat \Sigma_y^{\top}$ as formulated in Section~\ref{sec4}. Use $\hat q^{TVACLE}$  and $\hat q^{LWY}$ in \cite{li2017identifying} to determine the number of factors. All parameters in these two methods share the same settings.
We can see that the two largest eigenvalues of $\hat {\mathbf{M}}_y$ are $7.17\times 10^{-7}$ and $2.01\times 10^{-7}$, and the third to the $40$-th eigenvalues are shown in Figure~\ref{figure2}.

\begin{figure}[h]
\centering
\caption{\footnotesize Ratios $\hat\lambda_{i+1}/\hat\lambda_i$ in Li et al. (2017) and Ratios $\hat r_i^{TR}$ in the TVACLE, $1\le i\le 40$.}
\includegraphics[scale=0.5]{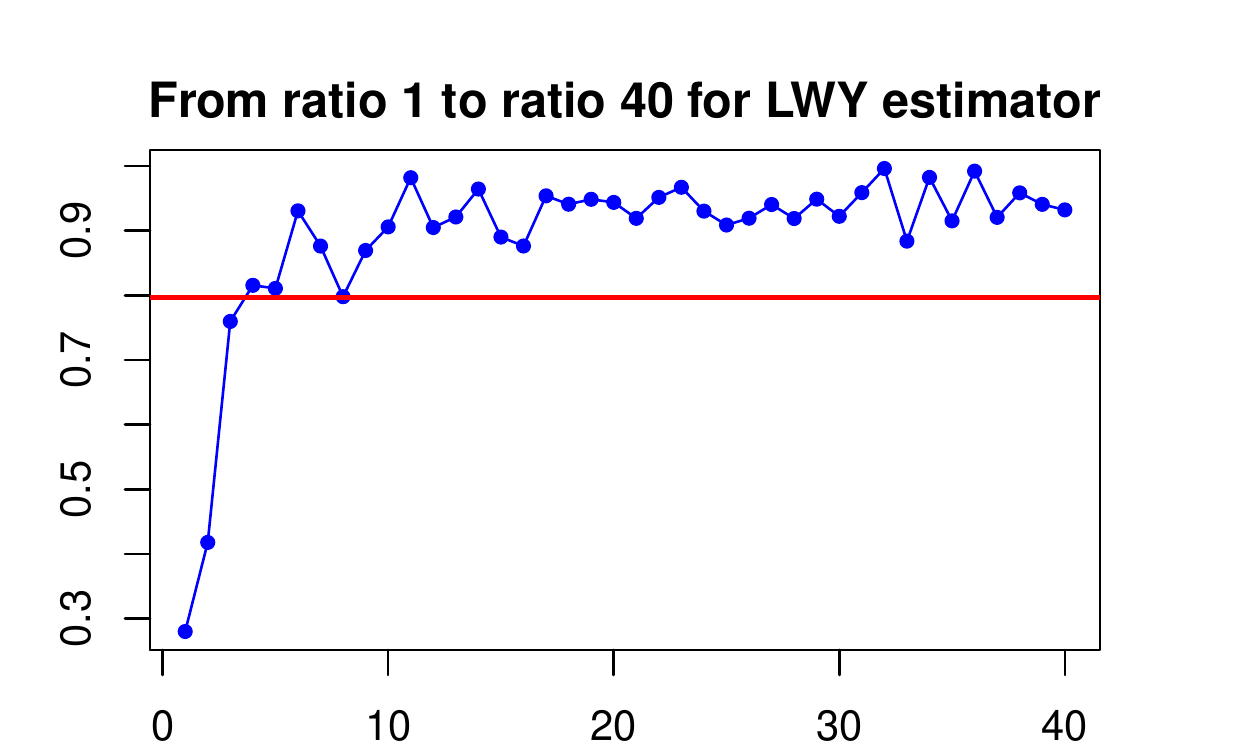}
\includegraphics[scale=0.5]{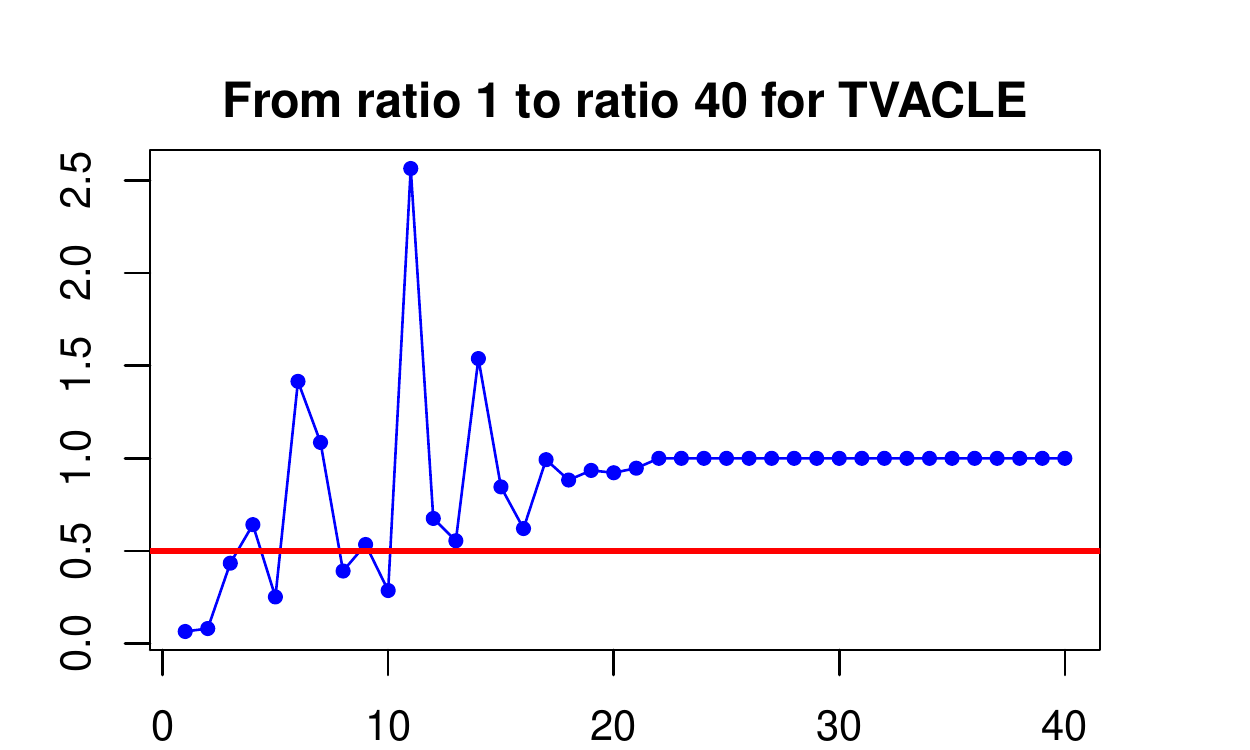}
\label{figure3}
\end{figure}

%Recall that  the estimation criterion in (\ref{LWY estimator}) proposed by \cite{li2017identifying}  defines $\hat q^{LWY}$ as the minimal index for which the next two consecutive ratios lie above a prefixed threshold ``$y=1-d_T$''. 
Figure~\ref{figure3} shows that $\hat q^{LWY}=5$. However, as shown in Figure~\ref{figure2}, the gap between the 5th eigenvalue and several following eigenvalues is evidently insignificant. %Hence, the estimator $\hat q^{LWY}$ has high risk of underestimation. A major reason behind this phenomenon comes from its construction of ``ratios".
As $\hat q^{LWY}$ is based on the magnitudes of the next two consecutive ratios. If eigenvalue multiplicity occurs, $\hat q^{LWY}$ could likely select a value smaller than the true number.

When the TVACLE is used, $\hat q^{TVACLE}=10$. %The first 40 ratios and the thresholding line $\tau=0.5$ are displayed in 
Figure~\ref{figure3} shows that the $11$th ratio is much larger than the $10$th ratio, although some values get smaller. Note that in this example, $c\sim 0.2$ and the ridge is relatively small, which would not very much dominate the difference between the eigenvalues and thus some oscillating values remain after the 10th ratio.

%The TVACLE identifies five more factors than $\hat q^{LWY}$. 
It is considered that $\hat q^{LWY}$ would neglect several factors and likely result in an underestimation. For a real data example, we usually cannot give a definitive answer. However, our method could provide an estimation that would be relatively conservative but necessary, particularly in the initial stage of data analysis; otherwise, an excessively parsimonious model would cause misleading conclusions.

\section{Concluding remarks}\label{sec7}
In this paper, we propose a valley-cliff criterion for spiked models, and the method can be applied to other order determination problems when the dimension is proportional to the sample size, such as those in sufficient dimension reduction if the corresponding asymptotics can be well investigated. The method is for the case with a fixed order $q$. An extension to the case with diverging $q$ will be proposed in our future work.

%%%%%%%%%%%%%%%%%%%%%%%%%%%%%%%%%%%%%%%%%%%%%%%%%%%%%%%%%%%%%%%%%%%%%%%%%%%%%%%%%%%%%%%%%%%%%%%%%%%%%%%%%%%%%%%%%%%%%%%%%%%%
\vskip 14pt
\noindent {\large\bf Supplementary Materials}

\noindent Proofs and  technical details are contained in the supplementary materials.
\par

%\bibliographystyle{Chicago}
%\bibliography{Ref}

\begin{thebibliography}{}

\bibitem[\protect\citeauthoryear{Baik, Arous, and P{\'e}ch{\'e}}{Baik
  et~al.}{2005}]{baik2005phase}
Baik, J., G.~B. Arous, and S.~P{\'e}ch{\'e} (2005).
\newblock Phase transition of the largest eigenvalue for nonnull complex sample
  covariance matrices.
\newblock {\em The Annals of Probability\/}~{\em 33\/}(5), 1643--1697.

\bibitem[\protect\citeauthoryear{Baik and Silverstein}{Baik and
  Silverstein}{2006}]{baik2006eigenvalues}
Baik, J. and J.~W. Silverstein (2006).
\newblock Eigenvalues of large sample covariance matrices of spiked population
  models.
\newblock {\em Journal of Multivariate Analysis\/}~{\em 97\/}(6), 1382--1408.

\bibitem[\protect\citeauthoryear{Bao, Pan, and Zhou}{Bao
  et~al.}{2015}]{bao2015universality}
Bao, Z., G.~Pan, and W.~Zhou (2015).
\newblock Universality for the largest eigenvalue of sample covariance matrices
  with general population.
\newblock {\em The Annals of Statistics\/}~{\em 43\/}(1), 382--421.

\bibitem[\protect\citeauthoryear{Benaych-Georges, Guionnet, and
  Maida}{Benaych-Georges et~al.}{2011}]{benaych2011fluctuations}
Benaych-Georges, F., A.~Guionnet, and M.~Maida (2011).
\newblock Fluctuations of the extreme eigenvalues of finite rank deformations
  of random matrices.
\newblock {\em Electronic Journal of Probability\/}~{\em 16\/}(60), 1621--1662.

\bibitem[\protect\citeauthoryear{Benaych-Georges and
  Nadakuditi}{Benaych-Georges and Nadakuditi}{2011}]{benaych2011eigenvalues}
Benaych-Georges, F. and R.~R. Nadakuditi (2011).
\newblock The eigenvalues and eigenvectors of finite, low rank perturbations of
  large random matrices.
\newblock {\em Advances in Mathematics\/}~{\em 227\/}(1), 494--521.

\bibitem[\protect\citeauthoryear{Han, Pan, and Yang}{Han
  et~al.}{2018}]{han2018unified}
Han, X., G.~Pan, and Q.~Yang (2018).
\newblock A unified matrix model including both {CCA} and {F} matrices in
  multivariate analysis: the largest eigenvalue and its applications.
\newblock {\em Bernoulli\/}~{\em 24\/}(4B), 3447--3468.

\bibitem[\protect\citeauthoryear{Han, Pan, and Zhang}{Han
  et~al.}{2016}]{han2016tracy}
Han, X., G.~Pan, and B.~Zhang (2016).
\newblock The tracy--widom law for the largest eigenvalue of {F} type matrices.
\newblock {\em The Annals of Statistics\/}~{\em 44\/}(4), 1564--1592.

\bibitem[\protect\citeauthoryear{Johnstone}{Johnstone}{2001}]{johnstone2001distribution}
Johnstone, I.~M. (2001).
\newblock On the distribution of the largest eigenvalue in principal components
  analysis.
\newblock {\em The Annals of Statistics\/}~{\em 29\/}(2), 295--327.

\bibitem[\protect\citeauthoryear{Johnstone and Lu}{Johnstone and
  Lu}{2009}]{johnstone2009consistency}
Johnstone, I.~M. and A.~Y. Lu (2009).
\newblock On consistency and sparsity for principal components analysis in high
  dimensions.
\newblock {\em Journal of the American Statistical Association\/}~{\em
  104\/}(486), 682--693.

\bibitem[\protect\citeauthoryear{Knowles and Yin}{Knowles and
  Yin}{2017}]{knowles2017anisotropic}
Knowles, A. and J.~Yin (2017).
\newblock Anisotropic local laws for random matrices.
\newblock {\em Probability Theory and Related Fields\/}~{\em 169\/}(1-2),
  257--352.

\bibitem[\protect\citeauthoryear{Kritchman and Nadler}{Kritchman and
  Nadler}{2008}]{kritchman2008determining}
Kritchman, S. and B.~Nadler (2008).
\newblock Determining the number of components in a factor model from limited
  noisy data.
\newblock {\em Chemometrics and Intelligent Laboratory Systems\/}~{\em
  94\/}(1), 19--32.

\bibitem[\protect\citeauthoryear{Lam and Yao}{Lam and
  Yao}{2012}]{lam2012factor}
Lam, C. and Q.~Yao (2012).
\newblock Factor modeling for high-dimensional time series: inference for the
  number of factors.
\newblock {\em The Annals of Statistics\/}~{\em 40\/}(2), 694--726.

\bibitem[\protect\citeauthoryear{Li}{Li}{1991}]{li1991sliced}
Li, K.-C. (1991).
\newblock Sliced inverse regression for dimension reduction.
\newblock {\em Journal of the American Statistical Association\/}~{\em
  86\/}(414), 316--327.

\bibitem[\protect\citeauthoryear{Li, Wang, and Yao}{Li
  et~al.}{2017}]{li2017identifying}
Li, Z., Q.~Wang, and J.~Yao (2017).
\newblock Identifying the number of factors from singular values of a large
  sample auto-covariance matrix.
\newblock {\em The Annals of Statistics\/}~{\em 45\/}(1), 257--288.

\bibitem[\protect\citeauthoryear{Luo and Li}{Luo and
  Li}{2016}]{luo2016combining}
Luo, W. and B.~Li (2016).
\newblock Combining eigenvalues and variation of eigenvectors for order
  determination.
\newblock {\em Biometrika\/}~{\em 103\/}(4), 875--887.

\bibitem[\protect\citeauthoryear{Nadler}{Nadler}{2010}]{nadler2010nonparametric}
Nadler, B. (2010).
\newblock Nonparametric detection of signals by information theoretic criteria:
  performance analysis and an improved estimator.
\newblock {\em IEEE Transactions on Signal Processing\/}~{\em 58\/}(5),
  2746--2756.

\bibitem[\protect\citeauthoryear{Onatski}{Onatski}{2009}]{onatski2009testing}
Onatski, A. (2009).
\newblock Testing hypotheses about the number of factors in large factor
  models.
\newblock {\em Econometrica\/}~{\em 77\/}(5), 1447--1479.

\bibitem[\protect\citeauthoryear{Passemier, Li, and Yao}{Passemier
  et~al.}{2017}]{passemier2017estimation}
Passemier, D., Z.~Li, and J.~Yao (2017).
\newblock On estimation of the noise variance in high dimensional probabilistic
  principal component analysis.
\newblock {\em Journal of the Royal Statistical Society: Series B (Statistical
  Methodology)\/}~{\em 79\/}(1), 51--67.

\bibitem[\protect\citeauthoryear{Passemier and Yao}{Passemier and
  Yao}{2012}]{passemier2012determining}
Passemier, D. and J.~Yao (2012).
\newblock On determining the number of spikes in a high-dimensional spiked
  population model.
\newblock {\em Random Matrices: Theory and Applications\/}~{\em 1\/}(01),
  1150002.

\bibitem[\protect\citeauthoryear{Passemier and Yao}{Passemier and
  Yao}{2014}]{passemier2014estimation}
Passemier, D. and J.~Yao (2014).
\newblock Estimation of the number of spikes, possibly equal, in the
  high-dimensional case.
\newblock {\em Journal of Multivariate Analysis\/}~{\em 127}, 173--183.

\bibitem[\protect\citeauthoryear{P{\'e}ch{\'e}}{P{\'e}ch{\'e}}{2006}]{peche2006largest}
P{\'e}ch{\'e}, S. (2006).
\newblock The largest eigenvalue of small rank perturbations of hermitian
  random matrices.
\newblock {\em Probability Theory and Related Fields\/}~{\em 134\/}(1),
  127--173.

\bibitem[\protect\citeauthoryear{Pillai and Yin}{Pillai and
  Yin}{2014}]{pillai2014universality}
Pillai, N.~S. and J.~Yin (2014).
\newblock Universality of covariance matrices.
\newblock {\em The Annals of Applied Probability\/}~{\em 24\/}(01), 935--1001.

\bibitem[\protect\citeauthoryear{Ulfarsson and Solo}{Ulfarsson and
  Solo}{2008}]{ulfarsson2008dimension}
Ulfarsson, M.~O. and V.~Solo (2008).
\newblock Dimension estimation in noisy pca with sure and random matrix theory.
\newblock {\em IEEE Transactions on Signal Processing\/}~{\em 56\/}(12),
  5804--5816.

\bibitem[\protect\citeauthoryear{Wang and Yao}{Wang and
  Yao}{2017}]{wang2017extreme}
Wang, Q. and J.~Yao (2017).
\newblock Extreme eigenvalues of large-dimensional spiked fisher matrices with
  application.
\newblock {\em The Annals of Statistics\/}~{\em 45\/}(1), 415--460.

\bibitem[\protect\citeauthoryear{Xia, Xu, and Zhu}{Xia
  et~al.}{2015}]{xia2015consistently}
Xia, Q., W.~Xu, and L.~Zhu (2015).
\newblock Consistently determining the number of factors in multivariate
  volatility modelling.
\newblock {\em Statistica Sinica\/}~{\em 25\/}(3), 1025--1044.

\bibitem[\protect\citeauthoryear{Zhu, Miao, and Peng}{Zhu
  et~al.}{2006}]{zhu2006sliced}
Zhu, L., B.~Miao, and H.~Peng (2006).
\newblock On sliced inverse regression with high-dimensional covariates.
\newblock {\em Journal of the American Statistical Association\/}~{\em
  101\/}(474), 630--643.

\bibitem[\protect\citeauthoryear{Zhu, Wang, Zhu, and Ferr{\'e}}{Zhu
  et~al.}{2010}]{zhu2010sufficient}
Zhu, L., T.~Wang, L.~Zhu, and L.~Ferr{\'e} (2010).
\newblock Sufficient dimension reduction through discretization-expectation
  estimation.
\newblock {\em Biometrika\/}~{\em 97\/}(2), 295--304.

\end{thebibliography}

\begin{thebibliography}{}

\bibitem[\protect\citeauthoryear{Bai and Yao}{Bai and
  Yao}{2008}]{bai2008central}
Bai, Z. and J.~Yao (2008).
\newblock Central limit theorems for eigenvalues in a spiked population model.
\newblock In {\em Annales de l'Institut Henri Poincar{\'e}, Probabilit{\'e}s et
  Statistiques}, Volume~44, pp.\  447--474. Institut Henri Poincar{\'e}.

\bibitem[\protect\citeauthoryear{Baik and Silverstein}{Baik and
  Silverstein}{2006}]{baik2006eigenvalues}
Baik, J. and J.~W. Silverstein (2006).
\newblock Eigenvalues of large sample covariance matrices of spiked population
  models.
\newblock {\em Journal of Multivariate Analysis\/}~{\em 97\/}(6), 1382--1408.

\bibitem[\protect\citeauthoryear{Bao, Pan, and Zhou}{Bao
  et~al.}{2015}]{bao2015universality}
Bao, Z., G.~Pan, and W.~Zhou (2015).
\newblock Universality for the largest eigenvalue of sample covariance matrices
  with general population.
\newblock {\em The Annals of Statistics\/}~{\em 43\/}(1), 382--421.

\bibitem[\protect\citeauthoryear{Benaych-Georges, Guionnet, and
  Maida}{Benaych-Georges et~al.}{2011}]{benaych2011fluctuations}
Benaych-Georges, F., A.~Guionnet, and M.~Maida (2011).
\newblock Fluctuations of the extreme eigenvalues of finite rank deformations
  of random matrices.
\newblock {\em Electronic Journal of Probability\/}~{\em 16\/}(60), 1621--1662.

\bibitem[\protect\citeauthoryear{Han, Pan, and Yang}{Han
  et~al.}{2018}]{han2018unified}
Han, X., G.~Pan, and Q.~Yang (2018).
\newblock A unified matrix model including both {CCA} and {F} matrices in
  multivariate analysis: the largest eigenvalue and its applications.
\newblock {\em Bernoulli\/}~{\em 24\/}(4B), 3447--3468.

\bibitem[\protect\citeauthoryear{Han, Pan, and Zhang}{Han
  et~al.}{2016}]{han2016tracy}
Han, X., G.~Pan, and B.~Zhang (2016).
\newblock The tracy--widom law for the largest eigenvalue of {F} type matrices.
\newblock {\em The Annals of Statistics\/}~{\em 44\/}(4), 1564--1592.

\bibitem[\protect\citeauthoryear{Li, Pan, and Yao}{Li
  et~al.}{2015}]{li2015singular}
Li, Z., G.~Pan, and J.~Yao (2015).
\newblock On singular value distribution of large-dimensional autocovariance
  matrices.
\newblock {\em Journal of Multivariate Analysis\/}~{\em 137}, 119--140.

\bibitem[\protect\citeauthoryear{Li, Wang, and Yao}{Li
  et~al.}{2017}]{li2017identifying}
Li, Z., Q.~Wang, and J.~Yao (2017).
\newblock Identifying the number of factors from singular values of a large
  sample auto-covariance matrix.
\newblock {\em The Annals of Statistics\/}~{\em 45\/}(1), 257--288.

\bibitem[\protect\citeauthoryear{Passemier and Yao}{Passemier and
  Yao}{2012}]{passemier2012determining}
Passemier, D. and J.~Yao (2012).
\newblock On determining the number of spikes in a high-dimensional spiked
  population model.
\newblock {\em Random Matrices: Theory and Applications\/}~{\em 1\/}(01),
  1150002.

\bibitem[\protect\citeauthoryear{Passemier and Yao}{Passemier and
  Yao}{2014}]{passemier2014estimation}
Passemier, D. and J.~Yao (2014).
\newblock Estimation of the number of spikes, possibly equal, in the
  high-dimensional case.
\newblock {\em Journal of Multivariate Analysis\/}~{\em 127}, 173--183.

\bibitem[\protect\citeauthoryear{Wang and Yao}{Wang and
  Yao}{2016}]{wang2016moment}
Wang, Q. and J.~Yao (2016).
\newblock Moment approach for singular values distribution of a large
  auto-covariance matrix.
\newblock In {\em Annales de l'Institut Henri Poincar{\'e}, Probabilit{\'e}s et
  Statistiques}, Volume~52, pp.\  1641--1666. Institut Henri Poincar{\'e}.

\bibitem[\protect\citeauthoryear{Wang and Yao}{Wang and
  Yao}{2017}]{wang2017extreme}
Wang, Q. and J.~Yao (2017).
\newblock Extreme eigenvalues of large-dimensional spiked fisher matrices with
  application.
\newblock {\em The Annals of Statistics\/}~{\em 45\/}(1), 415--460.

\end{thebibliography}

%%%%%%%%%%%%%%%%%%%%%%%%%%%%%%%%%%%%%%%%%%%%%%%%%%%%%%%%%%%%%%%%%%%%%%%%%%%%%%%%%%%%%%%%%%%%%%%%%%%%%%%%%%%%%%%%%%%%%%%%%%%%
\vskip .65cm
\noindent
Yicheng Zeng, Department of Mathematics, Hong Kong Baptist University, Hong Kong
\vskip 2pt
\noindent
E-mail: statzyc@163.com
\vskip 2pt
\noindent
Lixing Zhu, Department of Mathematics, Hong Kong Baptist University, Hong Kong\vskip 2pt
\noindent
E-mail: lzhu@hkbu.edu.hk

\newpage

\begin{center}
{\bf \large Supplementary Materials for  ``Order Determination for Spiked Models''}\\
{ Yicheng Zeng
    and
    Lixing Zhu\\ \vspace{-.2cm}
    Department of Mathematics, Hong Kong Baptist University, Hong Kong}
\end{center}
  
\spacingset{1.45} % DON'T change the spacing!

\hspace{.8cm}

\noindent
{\bf Proof of Theorem~2.1}
 We only need to check that
\begin{equation}
\lim_{n\rightarrow+\infty}\mathbb{P}\left(\dfrac{\hat\delta_{i+1}+\sigma^2 c_n}{\hat\delta_{i}+\sigma^2 c_n}>\tau\right)=1,\ for\ q<i\leq L-2,
\end{equation}
and
\begin{equation}
\lim_{n\rightarrow+\infty}\mathbb{P}\left(\dfrac{\hat\delta_{q+1}+\sigma^2 c_n}{\hat\delta_{q}+\sigma^2 c_n}\le\tau\right)=1.
\end{equation}
For any $q<i\leq L-2$, we have in probability
\begin{equation}
\dfrac{\hat\delta_{i+1}+\sigma^2 c_n}{\hat\delta_{i}+\sigma^2 c_n}=\dfrac{\hat\delta_{i+1}/\sigma^2+c_n}{\hat\delta_{i}/\sigma^2+c_n}\rightarrow 1>\tau,
\end{equation}
as $\hat\delta_{i+1}/\sigma^2=O_p(\tilde c_n)$ and $\tilde c_n=o(c_n)$. Thus, $\mathbb{P}(\frac{\hat\delta_{i+1}+\sigma^2 c_n}{\hat\delta_{i}+\sigma^2 c_n}>\tau)\to 1$, for $q<i\leq L-2$.
Further, in probability
\begin{equation}
\dfrac{\hat\delta_{q+1}+\sigma^2 c_n}{\hat\delta_{q}+\sigma^2 c_n}
=\dfrac{\hat\delta_{q+1}/\sigma^2+ c_n}{\hat\delta_{q}/\sigma^2+ c_n}
=\dfrac{o_p(1)+c_n}{d-e+o_p(1)+c_n}\rightarrow 0<\tau,
\end{equation}
as  $d-e>0$. Thus, $\mathbb{P}\left(\frac{\hat\delta_{q+1}+\sigma^2 c_n}{\hat\delta_{q}+\sigma^2 c_n}\leq \tau\right)\rightarrow1$. Therefore, the result is proved. \hfill$\Box$\\

\noindent{\bf Proof of Lemma~3.1}. Firstly, we check the requirement $(i)$.
As $f_n$ is differentiable, $\exists \xi_{q}\in (\hat\lambda_{q+1}/\sigma^2,\hat\lambda_{q}/\sigma^2),\ s.t.$
\begin{equation}
\hat\delta_{q}^*=f_n(\hat\lambda_{q}/\sigma^2)-f_n(\hat\lambda_{q+1}/\sigma^2)=f_n'(\xi_{q})\hat\delta_{q}.
\end{equation}
We then only need to check that
\begin{equation}
\mathbb{P}\{f_n'(\xi_{q})\geq1\}\longrightarrow1,\quad as\ n\rightarrow\infty.
\end{equation}
By conditions $(b)$ and $(c)$, it suffices to show that
\begin{equation}
\mathbb{P}\{\xi_{q}>e-\kappa_n\}\longrightarrow1,\quad as\ n\rightarrow\infty.
\end{equation}
On the other hand, from the definition of $\kappa_n$ in condition (c), we have
\begin{equation}
\hat\lambda_{q+1}/\sigma^2-e=o_P(\kappa_n),
\end{equation}
which is equivalent to
\begin{equation}
\dfrac{\hat\lambda_{q+1}/\sigma^2-e}{\kappa_n}=o_P(1) .
\end{equation}
Then we have
\begin{eqnarray}
\mathbb{P}\{\xi_{q}>e-\kappa_n\}\geq \mathbb{P}\left\{\frac{\hat\lambda_{q+1}}{\sigma^2}>e-\kappa_n\right\} =\mathbb{P}\left\{\dfrac{\hat\lambda_{q+1}/\sigma^2-e}{\kappa_n}>-1\right\}
\rightarrow 1.
\end{eqnarray}
 $(i)$ is then verified.\\
Now we check $(ii)$. Similarly, we have
\begin{eqnarray}
\hat\delta_{i}^*=f'(\xi_{i})\hat\delta_{i}/\sigma^2,\quad for\ q+1\leq i\leq p-2,
\end{eqnarray}
where $\xi_{i}\in(\hat\lambda_{i+1}/\sigma^2,\hat\lambda_{i}/\sigma^2)$. Then it suffices to show that
\begin{eqnarray}
\mathbb{P}\{\xi_{i}<e+\kappa_n\}\longrightarrow1,\quad for\ q+1\leq i\leq p-2.
\end{eqnarray}
Since $\xi_{q+1}>\cdots>\xi_{p-2}$, it is equivalent to
\begin{eqnarray}
\mathbb{P}\{\xi_{q+1}<e+\kappa_n\}\longrightarrow 1,
\end{eqnarray}
whose proof is completely parallel to that of $(i)$.\\
For $(iii)$, we have
\begin{eqnarray}
\dfrac{\hat\delta_{q+1}^*}{\hat\delta_{q}^*}=\dfrac{f_n'(\xi_{q+1})\hat\delta_{q+1}}{f_n'(\xi_{q})\hat\delta_{q}}
\end{eqnarray}
Condition $(b)$ yields
\begin{eqnarray}
f_n'(\xi_{q+1})\leq f_n'(\xi_{q}),
\end{eqnarray}
since $\xi_{q+1}<\hat\lambda_{q+1}/\sigma^2<\xi_{q}$.\ Therefore,
\begin{eqnarray}
\dfrac{\hat\delta_{q+1}^*}{\hat\delta_{q}^*}\leq \dfrac{\hat\delta_{q+1}}{\hat\delta_{q}}.
\end{eqnarray}
The requirement (iii) is then proved and the proof of the lemma is finished. \hfill$\Box$\\

\noindent{\bf Proof of Theorem~3.1}. Similar to the proof of Theorem~2.1, we only need to check that
\begin{eqnarray}
\lim_{n\to \infty}\mathbb{P}\left\{\dfrac{\hat\delta_{i+1}^*+c_n}{\hat\delta_{i}^*+c_n}>\tau\right\}=1,\quad for\ q<i\leq L-2
\end{eqnarray}
and
\begin{eqnarray}
\lim_{n\to \infty}\mathbb{P}\left\{\dfrac{\hat\delta_{q+1}^*+c_n}{\hat\delta_{q}^*+c_n}\leq \tau\right\}=1.
\end{eqnarray}
On one hand, since Lemma~3.1 ensures the requirement $(ii)$, for $q<i\leq L-2$,
\begin{eqnarray}
\hat\delta_{i}^*=o_p(c_n),
\end{eqnarray}
which leads to $\hat\delta_{i}^* c_n^{-1}=o_p(1)$.
Then
\begin{eqnarray}
\dfrac{\hat\delta_{i+1}^*+c_n}{\hat\delta_{i}^*+c_n}=\dfrac{\hat\delta_{i+1}^* c_n^{-1}+1}{\hat\delta_{i}^* c_n^{-1}+1}=\dfrac{o_p(1)+1}{o_p(1)+1}\xrightarrow{P}1>\tau.
\end{eqnarray}
That is,
\begin{eqnarray}
\lim_{n\to+\infty}\mathbb{P}\left\{\dfrac{\hat\delta_{i+1}^*+c_n}{\hat\delta_{i}^*+c_n}>\tau\right\}=1,\quad for\ q<i\leq L-2.
\end{eqnarray}
On the other hand, because of $(i)$, $(ii)$ and
\begin{eqnarray}
\lim_{n\to+\infty}\mathbb{P}\left\{\dfrac{\hat\delta_{q+1}/\sigma^2+c_n}{\hat\delta_{q}/\sigma^2+c_n}\leq\tau\right\}\longrightarrow1,
\end{eqnarray}
we have
\begin{eqnarray}
\lim_{n\to+\infty}\mathbb{P}\left\{\dfrac{\hat\delta_{q+1}^*+c_n}{\hat\delta_{q}^*+c_n}\leq\tau\right\}\longrightarrow1.
\end{eqnarray}
Thus, $\hat{q}_n^{TVACLE}$ is equal to $q$ with a probability going to $1$. The proof is concluded. \hfill$\Box$\\

\noindent{\bf Proof of Theorem~4.1.} \, First, we show that $U(F)=\sigma^2(1+\sqrt c)$ is the optimal lower bound we defines in Theorem~2.1.
    As \cite{passemier2012determining} and \cite{passemier2014estimation} pointed out, together with the results with $\sigma^2=1$ in \cite{baik2006eigenvalues}, a scale transformation $\hat\lambda_{i}\rightarrowtail (\sigma^2)^{-1} \hat\lambda_{i}$ can derive the results with  the general $\sigma^2$ as follows. When $0\le c\le 1$,  the empirical distribution of all estimated eigenvalues $\hat{\lambda}_{i}$
 almost surely converges to the famous Marcenko-Pastur distribution in the  support interval $(\sigma^2(1-\sqrt c)^2,\sigma^2(1+\sqrt c)^2):=(a,b)$. To be specific,
\begin{equation}\label{5}
\dfrac{1}{p}\#\{\hat{\lambda}_{i}:\hat{\lambda}_{i}<x\}\rightarrow F_{c,\sigma^2}(x)\quad a.s.
\end{equation}
with the density function
\begin{equation}\label{6}
F_{c,\sigma^2}'(x)=\dfrac{1}{2\pi xc\sigma^2}\sqrt{(b-x)(x-a)},\ a<x<b.
\end{equation}
 When $c>1$,   the integral of the above density function over the interval $(a, b)$ is equal to $1/c$, and there is an additional Dirac measure of mass $1-\frac{1}{c}$ at the origin $x=0$. These results show that completely unlike the case with the fixed $p$, there are the number of the estimated eigenvalues proportional to  $n$ larger than $\sigma^2$ and thus, using estimated eigenvalues to identify $q_1$  is impossible. Slightly generalizing the results of  \cite{baik2006eigenvalues}, we can  also have the phase transition phenomenon for spiked population models: for any fixed $L$ with $q+3 \le L <p$, %that $\hat\lambda_{n,i}$ converges to the upper bound $b$ of the support when its corresponding $\lambda_i$ lies below a threshold $1+\sqrt c$ but converges to the value $\phi (\lambda_i):=\lambda_i+\frac{c\lambda_i}{\lambda_i-1}>b$ when $\lambda_i$ is above the threshold. That is,
\begin{eqnarray}\label{8}
\hat \lambda_{i}&\rightarrow & \sigma^2\phi (\lambda_i/\sigma^2)\ a.s.\ \forall\ i\leq q ,\\
%\end{equation}
%\begin{equation}\label{9}
\hat \lambda_{i}&\rightarrow & b\quad a.s., \ for \ q+1 \le i\leq L ,
\end{eqnarray}
where $q:=\# \{ \lambda_i:\lambda_i>\sigma^2(1+\sqrt c) \}$, and $\phi(x):=x+\frac{cx}{x-1}$ which is a strictly increasing function on $(1+\sqrt c,+\infty )$.  %In the following, without confusion, we rewrite $\lambda_{n,i}$ for $1\leq i\leq q_1$  as $\lambda_i$ throughout the best of this paper.
That is to say, only $q$ extreme sample eigenvalues would converge to values larger than the right hand end $b$ of the interval $(a,b)$ if their corresoponding spikes exceed the value $\sigma^2(1+\sqrt c)$. Otherwise, the estiated eigenvalues corresponding to $\sigma^2 < \lambda_i \le \sigma^2(1+\sqrt c)$ for $ q+1 \le i\leq L$ converge to the same value $b$. This causes that these estimated eigenvalues are inseparable from those of $\lambda_i=\sigma^2$.

Further,
\cite{bai2008central} established  Central Limit Theorem of $\hat\lambda_{i}$, for $1\leq i\leq q$ that implies the $\sqrt n$-consistency of $\hat\lambda_{i}$. Combining Proposition 5.8 of \cite{benaych2011fluctuations} and Corollary 1.5 and Remark 1.6 in \cite{bao2015universality}, we have that, for any fixed integer $L> q+1$, $\hat\lambda_{i}-b=O_P(n^{-2/3})$ and then $\hat\delta_{i}=O_P(n^{-2/3})$ for  $q+1\leq i\leq L$, where $b=\sigma^2(1+\sqrt c)^2$.
%Therefore, those spikes $\lambda_i$ not exceeding $\sigma^2(1+\sqrt c)$ is impossible to be separated from those non-spikes $\lambda_i=\sigma^2$.  Thus, for determining the number of spikes in this paper, we only consider the number $q \, (\le q_1)$ of spikes larger than $\sigma^2(1+\sqrt c)$.

According to the property of $\hat\lambda_{i}$, we have
\begin{equation}
\lim_{n\to\infty} \hat \delta_{i}= \begin{cases} \sigma^2\phi (\lambda_i/\sigma^2)-\sigma^2\phi (\lambda_{i+1}/\sigma^2)\ a.s. & for\ 1\leq i\leq q-1,\\
 \sigma^2\phi (\lambda_q/\sigma^2)-b>0\ a.s. & for \ i= q,\\
 0,\ a.s. & for \ q+1\leq i \leq L-1. \end{cases}
\end{equation}
%Precisely, $\hat{\delta}_{n,i}$'s associated with $\hat{\lambda}_{n,i}$ which are located inside the bulk converge to $0$ at a rate of  higher order than $O_P(n^{-2/3})$.
%Consider $a<x<x+\Delta x<b$. The formula (\ref{5}) implies that
%\begin{eqnarray*}
%\dfrac{1}{p}\#\{\hat \lambda_{n,i}:\hat \lambda_{n,i}\in[x,x+\Delta x)\}&\rightarrow&F_{c,\sigma^2}(x+\Delta x)-F_{c,\sigma^2}(x)\\
%&=&\int_x^{x+\Delta x}F'(t)dt\\
%&=&\int_x^{x+\Delta x}\dfrac{1}{2\pi tc}\sqrt{(b-t)(t-a)}dt\\
%&=&O(\Delta x).
%\end{eqnarray*}
%Thus, $\#\{\hat \lambda_{n,i}:\hat \lambda_{n,i}\in[x,x+\Delta x)\}=O(p\Delta x)$. Given a fixed $\Delta x$, the corresponding $\hat{\delta}_{n,i}=O_P(p^{-1})=O_P(n^{-1})=o_p(n^{-2/3})$ for all $i\in \{i:\hat \lambda_{n,i}\in [x,x+\Delta x)\}$. Hence, for any constant $0<\alpha<1$,
%\begin{eqnarray}
%\hat{\delta}_{n,i}=O_p(n^{-2/3}),\ for\ q+1\le i\le [\alpha \min\{n,p\}].
%\end{eqnarray}
Then the ratios have the following valley-cliff property at the index $q$: if defining $0/0$ as $1$,
\begin{eqnarray}
\lim_{n\to\infty}\hat r_{i}=\lim_{n\to\infty}\dfrac{\hat\delta_{i+1}}{\hat\delta_{i}}&=&\begin{cases}
\dfrac{\sigma^2\phi(\lambda_{i+1}/\sigma^2)-\sigma^2\phi(\lambda_{i+2}/\sigma^2)}{\sigma^2\phi(\lambda_i/\sigma^2)-\sigma^2\phi(\lambda_{i+1}/\sigma^2)}, & for\ 1\leq i\leq q-2 \nonumber\\
\dfrac{\sigma^2\phi(\lambda_{q}/\sigma^2)-b}{\sigma^2\phi(\lambda_{q-1}/\sigma^2)-\sigma^2\phi(\lambda_{q}/\sigma^2)}, & for\  i= q-1\\ 0,&for\ i=q\\0/0=1,&for\ q+1\leq i\leq L-2.\end{cases}\\
&=&\begin{cases}
\dfrac{\phi(\lambda_{i+1}/\sigma^2)-\phi(\lambda_{i+2}/\sigma^2)}{\phi(\lambda_i/\sigma^2)-\phi(\lambda_{i+1}/\sigma^2)}, & for\ 1\leq i\leq q-2\\
\dfrac{\phi(\lambda_{q}/\sigma^2)-(1+\sqrt c)^2}{\phi(\lambda_{q-1}/\sigma^2)-\phi(\lambda_{q}/\sigma^2)}, & for\  i= q-1\\ 0,&for\ i=q\\0/0=1,&for\ q+1\leq i\leq L-2.\end{cases}
\end{eqnarray}
From this result, we can now consider the ridge ratios.
Note that  $\hat\delta_{i}=O_P(n^{-2/3})$ for $q+1\le i\le L-1$ and $\hat\delta_{i}$ for $1\le q$, at the rate of order $O_P(n^{-1/2})$, are either consistent to positive constants or to $0$ when spikes are equal. Further, as $c_n\to 0$ and  $c_n^{-1}n^{-2/3}=o(1)$, we can easily see that
$\lim_{n\to+\infty}\hat r_{q}=0/( \phi (\lambda_q)-b)=0$ and then $\hat r_{q}^R\to 0$ in probability. We then still have the valley-cliff property at the index $q$:
\begin{equation}
\lim_{n\to\infty}\hat r_{i}^R=\begin{cases} \ge 0,&i<q\\0,&i=q\\1,&q+1\leq i\leq L-2. \end{cases}
\end{equation}
%The value of $\hat{r}_{n,i}^R$, for $i<q$, depends on the choice of ridge sequence $c_n$.
\hfill$\Box$\\

\noindent{\bf Proof of Theorem~4.2.} \,
When $\sigma^2=1$,  \cite{wang2017extreme} provided the limiting spectral distribution (LSD) of the matrix $\textbf{F}_n=\textbf{S}_1 \textbf{S}_2^{-1}$ and established the phase transition phenomenon for those extreme eigenvalues of $\textbf{F}_n$. When $0<c\leq 1$, the empirical spectral distribution (ESD) of $\textbf{F}_n$ weakly converges to a distribution $F_{c,y}$ with the density function
\begin{eqnarray}
f_{c,y}(x)=\frac{(1-y)\sqrt {(b_1-x)(x-a_1)}}{2\pi x(c+xy)},\quad a_2\leq x\leq b_2,
\end{eqnarray}
where $a_2=(\frac{1-\sqrt {c+y-cy}}{1-y})^2$ and $b_2=(\frac{1+\sqrt {c+y-cy}}{1-y})^2$. Similarly as that of spiked population models,  when $c>1$, there is an additional probability measure of mass $1-\frac{1}{c}$ for $F_{c,y}$. Further, they also proved a phase transition phenomenon that almost surely
\begin{eqnarray}
        \hat \lambda_{i} &\rightarrow&\varphi(\lambda_i),\quad \lambda_i>\gamma(1+\sqrt{c+y-cy}), \nonumber\\
        \hat \lambda_{i} &\rightarrow& b_2,\quad 1<\lambda_i\leq \gamma(1+\sqrt{c+y-cy}), \nonumber
 %       \hat \lambda_{n,i} &\rightarrow& a_2,\quad \gamma(1+\sqrt{c+y-cy})\leq\lambda_i<1,\nonumber
\end{eqnarray}
where $\gamma=\frac{1}{1-y}\in (1,+\infty)$ and $\varphi(x)=\frac{\gamma x(x-1+c)}{x-\gamma},\ x\neq \gamma$.

Under the general Fisher matrix with the spiked structure
 \begin{equation}
\text{spec}(\Sigma_1\Sigma_2^{-1})=\{\lambda_1,\lambda_2,\cdots,\lambda_{q_1},\sigma^2,\cdots,\sigma^2\}.
\end{equation}
Using the simple transformation $\hat\lambda_{i}\rightarrowtail (\sigma^2)^{-1}\hat\lambda_{i}$, we can similarly achieve the results in the case of $\sigma^2=1$. The empirical spectral distribution of $\textbf{F}_n$ weakly converges to a distribution $F_{c,y,\sigma^2}$ with the density function
\begin{equation}
f_{c,y,\sigma^2}(x)=\frac{1}{\sigma^2}f_{c,y}(\frac{x}{\sigma^2}),\quad \sigma^2a_1<x<\sigma^2b_1,
\end{equation}
and the additional point mass $1-\frac{1}{c}$ at origin $x=0$ also exists when $c>1$. The phase transition phenomenon is modified as
\begin{eqnarray}
        \hat \lambda_{i} &\rightarrow&\sigma^2\varphi(\lambda_i/\sigma^2),\quad \lambda_i>\sigma^2\gamma(1+\sqrt{c+y-cy}), \nonumber\\
        \hat \lambda_{i} &\rightarrow& \sigma^2b_2,\quad \sigma^2<\lambda_i\leq \sigma^2\gamma(1+\sqrt{c+y-cy}), \nonumber
 %       \hat \lambda_{n,i} &\rightarrow& \sigma^2a_2,\quad \sigma^2\gamma(1+\sqrt{c+y-cy})\leq\lambda_i<\sigma^2,
\end{eqnarray}
where the parameters  $b_2$, $\gamma$ and the function $\varphi$ have the same definitions as those in the case with $\sigma^2=1$.
%Here, we only consider those large $\lambda_i$'s in the sense that $\lambda_i>\sigma^2\gamma(1+\sqrt{c+y-cy})$.
Let $q:=\#\{\lambda_i:\lambda_i>U(F)=\sigma^2\gamma(1+\sqrt{c+y-cy})\}$. According to these results, for any fixed  $L$ with $q+3< L<p$
\begin{eqnarray}
        \hat \lambda_{i} &\rightarrow&\sigma^2\varphi(\lambda_i/\sigma^2),\quad 1\leq i\leq q, \nonumber\\
        \hat \lambda_{i} &\rightarrow& \sigma^2b_2,\quad q+1\leq i\leq L.
\end{eqnarray}
That is, when $i$ is larger than $q$, the estimated eigenvalue $\hat\lambda_{i}$  converges to the right edge $\sigma^2b_2$ of the support of $F_{c,y,\sigma^2}$. This means  that any eigenvalues such that $\sigma^2 <\lambda_i\le \sigma^2\gamma(1+\sqrt{c+y-cy})$ cannot be identified through the estimated eigenvalues and then show the optimality of this lower bound.
%So we can use the TVACLE criterion to conduct order determination for the matrix $F$, namely estimate $q$, when this transition phenomenon occurs for all $\lambda_i$'s.\\

 Modifying the result of \cite{wang2017extreme}, we can  show that  those extreme eigenvalues $\hat{\lambda}_{i}$ corresponding to $\lambda_i>\sigma^2\gamma(1+\sqrt{c+y-cy})$ satisfy  Central Limiting Theorem and thus have the convergence rate of order $1/\sqrt n$. For the fluctuation of those eigenvalues which stick to the bulk, \cite{han2016tracy} showed that $n^{2/3}(\hat{\lambda}_{q+1}-\sigma^2b_2)$ is asymptotically Tracy-Widom distributed. \cite{han2018unified} established an asymptotic joint distribution for $(n^{2/3}(\hat{\lambda}_{q+1}-\sigma^2b_2),n^{2/3}(\hat{\lambda}_{q+2}-\sigma^2b_2),\cdots,n^{2/3}(\hat{\lambda}_{q+k}-\sigma^2b_2))$ for any fixed $k$. Thus, for any fixed $L>q$, $n^{2/3}(\hat{\lambda}_{i}-\sigma^2b_2)=O_p(n^{-2/3})$ for $q+1\le i\le L$. The Spiked Fisher matrix $\textbf{F}_n$ satisfies  Model Feature~2.1. The proof is finished.
\hfill$\Box$\\

\noindent{\bf Proof of Proposition~4.2.}\ Let $\Sigma_y=cov(y_t,y_{t-1})$ be the lag-1 auto-covariance matrices of $y_t$ and $\Sigma_x=cov(x_t,x_{t-1})$ the log-1 auto-covariance matrix of $x_t$. As shown in \cite{li2017identifying}, the sample auto-covariance matrix of $y_t$ is
\begin{eqnarray}\label{auto}
\hat{\Sigma}_y&=&\frac{1}{T}\sum_{t=2}^{T+1}y_ty_{t-1}'=\frac{1}{T}\sum_{t=2}^{T+1}(\textbf{A}x_t+\varepsilon_t)(\textbf{A}x_{t-1}+\varepsilon_{t-1})' \nonumber\\
&=&\frac{1}{T}\sum_{t=2}^{T+1}\textbf{A}x_tx_{t-1}'\textbf{A}'+\frac{1}{T}\sum_{t=2}^{T+1}(\textbf{A}x_t\varepsilon_{t-1}'+\varepsilon_tx_{t-1}'\textbf{A}')+\frac{1}{T}\sum_{t=2}^{T+1}\varepsilon_t\varepsilon_{t-1}' \nonumber\\
&:=&\textbf{P}_\textbf{A}+\hat{\Sigma}_{\varepsilon},
\end{eqnarray}
where the matrix $\hat{\Sigma}_{\varepsilon}=\frac{1}{T}\sum_{t=2}^{T+1}\varepsilon_t\varepsilon_{t-1}'$ is the sample auto-covariance matrix of noise sequence $\{\varepsilon_t\}$. Notice that the matrix $\textbf{P}_\textbf{A}$ is of finite rank, then the matrix $\hat{\Sigma}_y$ can be viewed as a finite-rank perturbation of $\hat{\Sigma}_{\varepsilon}$. Since both $\hat{\Sigma}_{\varepsilon}$ and $\hat{\Sigma}_y$ are asymmetric matrices, \cite{li2017identifying} considered their singular values. This is equivalent to considering the square root of the eigenvalues of the matrices $\hat{\textbf{M}}_{\varepsilon}:=\hat{\Sigma}_{\varepsilon}\hat{\Sigma}_{\varepsilon}'$ and $\hat{\textbf{M}}_y:=\hat{\Sigma}_y\hat{\Sigma}_y'$, respectively.

Define $\hat \Sigma_y/\sigma^2=\textbf{P}_\textbf{A}/\sigma^2+\hat{\Sigma}_{\varepsilon}/\sigma^2$, we can reduce the problem to the case with $\sigma^2=1$.
When $p/T\rightarrow y>0$, \cite{li2015singular} proved that the empirical spectral distribution of $\hat{\textbf{M}}_{\varepsilon}$ almost surely converges to a non-random limiting distribution, whose Stieltjes transformation $\mathcal{S}(z)$ defined in (4.10) satisfies the equation
$$z^2\mathcal{S}^3(z)-2z(y-1)\mathcal{S}^2(z)+(y-1)^2\mathcal{S}(z)-z\mathcal{S}(z)-1=0.$$
This limiting spectral distribution is continuous with  a compact support $[a_1{\bf 1}_{\{y\ge 1\}},b_1]$, where $$a_1=(-1+20y+8y^2-(1+8y)^{3/2})/8,$$  $$b_1=(-1+20y+8y^2+(1+8y)^{3/2})/8$$
From \cite{wang2016moment}, the  largest eigenvalue $\hat \lambda_{\varepsilon, 1}$ of $\hat{\textbf{M}}_{\varepsilon}$ almost surely converges to the right edge $b_1$. Like the previous models, for any fixed $L>q_0+1$, and any $1\le i\le L$ the  largest eigenvalues $\hat \lambda_{\varepsilon, i}$ of $\hat{\textbf{M}}_{\varepsilon}$ converge to the same value $b_1$.
Further, for general $\sigma^2$,  the result of \cite{li2017identifying} implies that  the limiting spectral distribution of the perturbed matrix $\hat{\textbf{M}}_y$ is identical to that of $\hat {\textbf{M}}_{\varepsilon}$. They also built a phase transition phenomenon for those extreme eigenvalues $\hat \lambda_{1}\ge \cdots \ge \hat \lambda_{q}$. % as long as the variance $\gamma_0(i)$ and lag-1 auto-covariance $\gamma_1(i)$, $1\le i\le q_0$, of those factors satisfy certain condition.
%the strength of factors is strong enough.
The following proposition  confirms the optimality of the  bound restriction $\mathcal{T}_1(i)< \mathcal{T}(b_1+)$ such that the corresponding  $q$ factors in $\textbf{P}_\textbf{A}$ can be identified.\\

\begin{lemma}
[\cite{li2017identifying}]
Denote $\mathcal{T}(\cdot)$ as the $T$-transformation of the Limiting Spectral Distribution (LSD) for matrix $\hat {\mathbf{M}}_y/\sigma^4$. Suppose that the model (4.9) satisfies Assumptions~4.1 and 4.2, $\{\varepsilon_t\}$ are normally distributed and the loading matrix $\mathbf{A}$ is standardized as $\mathbf{A}'\mathbf{A}=\mathbf{I}_k$. Let $\hat \lambda_{i},\ 1\le i\le q_0$ denote the $q_0$ largest eigenvalues of $\hat {\mathbf{M}}_y$. Then for each $1\le i\le q_0$, $\hat \lambda_{i}/\sigma^4$ converges almost surely to a limit $\beta_i$. Moreover,
$$\beta_i>b_1
\, when \, \mathcal{T}_1(i)< \mathcal{T}(b_1+),$$
and
$$\beta_i=b_1 \,  when \,  \mathcal{T}_1(i)\ge \mathcal{T}(b_1+)$$
where
$$\mathcal{T}_1(i)=\frac{2y\sigma^2\gamma_0(i)+\gamma_1(i)^2-\sqrt{(2y\sigma^2\gamma_0(i)+\gamma_1(i)^2)^2-4y^2\sigma^4(\gamma_0(i)^2-\gamma_1(i))^2}}{2\gamma_0(i)^2-2\gamma_1(i)^2}.$$
%Otherwise, i.e. $\mathcal{T}_1(i)< \mathcal{T}(b_1+)$, then $\beta_i>b_1$.
%More specifically, the value of $\beta_i$ is determined by the equation:
%$$(y\sigma^2-\gamma_0(i)\mathcal{T}(\beta_i))^2=\gamma_1(i)^2\mathcal{T}(\beta_i)(1+\mathcal{T}(\beta_i)).$$
\end{lemma}

From this proposition, we can see that the bound for the number of common factors determined by the constraint $\mathcal{T}_1(i)< \mathcal{T}(b_1+)$ is optimal. That is,  only $q$ common factors in $\textbf{P}_\textbf{A}$ can be well separated from the noise $\varepsilon_t$'s theoretically. This is  because  $\hat \lambda_{q+1}$ will converge to $b_1$ and thus cannot be well separated from those large estimated eigenvalues of $\hat \Sigma_{\varepsilon}$ that tend to the right edge $b_1$ as well.  \hfill$\Box$\\

\noindent{\bf A justification of Proposition~4.2.} By the results of \cite{wang2016moment},  the condition (A1) holds. Further, under the assumption that  the estimated eigenvalues $\hat \lambda_i$ for $i>q$   have the convergence rate of order $O_p(n^{-2/3})$,  the condition (A2) in Model Feature~2.1  holds. Following the arguments used in Theorems~2.1 and 3.1, the results hold. \hfill$\Box$

\end{document}